\documentclass[twocolumn,english,prd]{revtex4}
\usepackage{graphicx}
\newcommand{\beq}{\begin{equation}}
\newcommand{\eeq}{\end{equation}}
\newcommand{\beqa}{\begin{eqnarray}}
\newcommand{\eeqa}{\end{eqnarray}}

\newcommand{\lexp}{\mathop{\langle}}
\newcommand{\rexp}{\mathop{\rangle}}

\def\d{\delta}
\def\te{\theta}
\def\ds{\delta_s}

\def\dD{\delta_{\rm D}}

\def\pl{L}

\font\BF=cmmib10

\def\k{{\hbox{\BF k}}}
\def\x{{\hbox{\BF x}}}
\def\r{{\hbox{\BF r}}}
\def\s{{\hbox{\BF s}}}

\def\p{{\hbox{\BF p}}}
\def\q{{\hbox{\BF q}}}
\def\v{{\hbox{\BF v}}}
\def\u{{\hbox{$u_z$}}}
\def\ub{{\hbox{\BF u}}}

\def\tvk{{\hat{\k}}}
\def\tvq{{\hat{\q}}}

\def\M{{\cal M}}

\def\rpa{r_{\parallel}}
\def\rpe{r_{\perp}}
\def\spa{s_{\parallel}}
\def\spe{s_{\perp}}

 \def\la{\mathrel{\mathpalette\fun <}}
 \def\ga{\mathrel{\mathpalette\fun >}}
 \def\fun#1#2{\lower3.6pt\vbox{\baselineskip0pt\lineskip.9pt
        \ialign{$\mathsurround=0pt#1\hfill##\hfil$\crcr#2\crcr\sim\crcr}}}

\begin{document}

%
%
\title{Redshift-Space Distortions, Pairwise Velocities and Nonlinearities}
%
%
\author{Rom\'an Scoccimarro}

\vskip 1pc

\address{Center for Cosmology and Particle Physics, \\ 
Department of Physics, New York University, \\
4 Washington Place, New York, NY 10003, USA}

\begin{abstract}
%

We derive the exact relationship, including all non-linearities, between real-space and redshift-space two-point statistics through the pairwise velocity distribution function. We show using numerical simulations that the pairwise velocity PDF is strongly non-Gaussian at all scales, and explain why this is so. We caution that a commonly used ansatz to model the redshift-space power spectrum gives rise to an unphysical distribution of  pairwise velocities, and show that it is in general impossible to derive the distribution from measurements of redshift-space clustering. Methods that claim to do this obtain instead something else, whose properties we derive. 

We provide a general derivation of the large-scale limit of the redshift-space power spectrum and show that it differs from the Kaiser formula by terms that depend on Gaussian and non-Gaussian contributions to the velocity dispersion of large-scale flows. We also show that the large-scale evolution of velocity fields is not well described by linear theory and discuss how this impacts the redshift-space power spectrum. Finally, we stress that using the monopole of the redshift-space power as an indicator of the real-space power spectrum shape can lead to systematic effects in the determination of cosmological parameters; nevertheless a simple procedure is able to recover the large-scale real-space power spectrum rather well.

\end{abstract}



\maketitle

%
%
\section{Introduction}
%
%

Redshift surveys provide a three-dimensional view of the large-scale structure of the universe. This view, however, is somewhat distorted due to gravitationally-induced peculiar velocities that contribute to galaxy redshifts in addition to the smooth Hubble flow. These ``redshift distortions" complicate the interpretation of galaxy clustering data from redshift surveys but, on the other hand, provide a measure of the amount of dark matter in the universe (which sources peculiar velocities) due to the induced anisotropy of clustering statistics such as the power spectrum and the two-point correlation function.

The two main signatures of peculiar velocities on the redshift-space clustering pattern have been known for a long time~\cite{1972MNRAS.156P...1J,1977ApJ...212L...3S,1980lssu.book.....P,1987MNRAS.227....1K}). At large scales, galaxies that fall into clusters look squashed along the line of sight in redshift space: infall velocities of galaxies between the cluster and us (between the cluster and the rest of the universe) add (substract) to the Hubble flow. This squashing effect leads to an increase of the clustering amplitude along the line of sight, thus the power spectrum is enhanced for waves parallel to the line of sight~\cite{1987MNRAS.227....1K}. At small scales (compared to the size of virialized clusters)  the internal velocity dispersion elongates clusters along the line of sight, leading to the so-called ``finger of god" effect. This suppresses the amplitude of waves parallel to the line of sight. Therefore,  the Fourier space clustering pattern shows a positive quadrupole anisotropy at large scales that gradually becomes smaller and eventually negative as small scales are probed~\footnote{In configuration space, the sign of the quadrupole is reversed from that in Fourier space; for a multipole moment of order $\ell$ they are related by a factor $i^\ell$ ($\ell$ must be  even by statistical isotropy), see Eq.~(\ref{Dsell}) below. At small scales it is easier to characterize multipoles in configuration space, they are positive for all $\ell$ and of comparable magnitude.}. 

This picture is captured by the ``dispersion model" for the redshift-space power spectrum,

\beq
P_s(k,\mu) = P_g(k) \ (1+\beta \mu^2)^2\ \frac{1}{1+k^2\mu^2\sigma_p^2/2},
\label{DMexp}
\eeq

where $P_g(k)$ is the real-space galaxy power spectrum, $\beta \approx \Omega_m^{0.6}/b_1$ where $b_1$ is the linear bias factor between galaxies and mass, $\mu = k_z/k$ with $\hat{z}$ denoting the line of sight direction, and $\sigma_p$ the pairwise velocity dispersion assumed to be a constant independent of scale. Here $\beta$ quantifies the squashing effect, $\sigma_p$ the velocity dispersion effect. The particular form for the squashing effect is due to linear dynamics and linearized real-to-redshift space mapping~\cite{1987MNRAS.227....1K}; hereafter {\em Kaiser limit}); the velocity dispersion factor is that corresponding to an exponential pairwise velocity distribution function with no mean streaming~\cite{1996MNRAS.282..877B}. These  effects factorize due to the implicit assumption that they can be treated as independent. Other dispersion models assume different dispersion factors (e.g.~\cite{1994MNRAS.267.1020P,1994ApJ...431..569P}).

The model in Eq.~(\ref{DMexp}) is clearly oversimplified for a number of reasons, among them 

\begin{enumerate}

\item[i)] Even in the context of linear dynamics from Gaussian initial conditions the squashing factor in Eq.~(\ref{DMexp}) must be an approximation. In a random Gaussian field, the velocity field fluctuates from point to point, so there is velocity dispersion and thus the squashing effect must be necessarily accompanied by some sort of dispersion effect. This implies that these effects are not independent.

\item[ii)] The dispersion factor introduces a phenomenological parameter $\sigma_p$ which represents an effective pairwise velocity dispersion, but whose value cannot be directly used to constrain models since in reality velocity dispersion is a function of scale and galaxy bias and is not clear how to relate it to the effective value $\sigma_p$ affecting the redshift-space power spectrum in this model.

\end{enumerate}

The first point implies that some of the dispersion effect may come from large-scale flows (as opposed to virial velocities) which can be modeled accurately in terms of the primordial power spectrum and cosmological parameters. This is important because such improvement  of the model can add significant constraining power on theories. The second point also implies that there is potentially a lot to be gained from finding exactly how the redshift-space power spectrum depends on non-linear effects from velocities and galaxy bias. Some attempts to do this using the halo model have been proposed~\cite{2001MNRAS.321....1W,2001MNRAS.325.1359S,2002MNRAS.336..892K}, but they do not address the first point made above. In addition, fitting formulae extracted from simulations that improve on the dispersion model have been developed~\cite{1999MNRAS.310.1137H,2001ApJ...547..545J}, which although very valuable, they do not provide much insight into the problem.

Despite its limitations, Eq.~(\ref{DMexp}) has been a popular model for analyzing redshift surveys to obtain constraints on cosmological parameters (e.g.~\cite{2001Natur.410..169P,2003ApJS..148..195V}). An alternative to using the dispersion model has been to simply ignore the dispersion effect, setting $\sigma_p=0$ in Eq.~(\ref{DMexp}), and argue that on ``large enough" scales this is sufficiently accurate. Many results on cosmological parameters from measurements of the power spectrum rest on this assumption (e.g.~\cite{2002MNRAS.333..961L,2002PhRvL..89f1301E,2002MNRAS.335..887T,2004ApJ...607..655P}). Although in the past uncertainties from redshift surveys have been large enough that such strategies were reasonable, present datasets such as 2dFGRS and SDSS demand better accuracy; moreover, one expects to get more information than just one or two numbers from using the full dependence of the redshift-space power spectrum on scale and direction. 

In this paper we derive an exact formula for the redshift-space two-point function and power spectrum in terms of the real space density and velocity fields, extending previous work along these lines~\cite{1999ApJ...517..531S}. We also show that this formula obeys a modified version of  the ``streaming model", which was previously proposed in the small-scale~\cite{1980lssu.book.....P} and large-scale~\cite{1995ApJ...448..494F} approximations. This gives a useful characterization of redshift distortions, since real and redshift space spectra are then related by the pairwise velocity probability distribution function (PDF), or its Fourier transform. The challenge is then how to model this PDF in terms of the linear power spectrum, cosmological parameters and galaxy bias. Some steps in this direction, modeling the first two moments, have been already given by~\cite{2001MNRAS.325.1288S,2001MNRAS.326..463S} using the halo model, see also~\cite{1998ApJ...504L...1J} for a modeling of the PDF inspired by perturbation theory. Recent work~\cite{TZW04} provides a modeling of the pairwise PDF starting from that of halos. In addition, we show that the model in Eq.~(\ref{DMexp}) leads to an unphysical distribution of pairwise velocities, and that inferring the pairwise velocity PDF from redshift-space clustering is unfortunately not possible in general. Methods that claim to do this~\cite{1998ApJ...494L.133L,2002ApJ...567L...1L,2003MNRAS.346...78H} recover instead something else, whose properties we derive here.

In this paper we mostly concentrate in the large-scale limit, showing using perturbation theory and N-body simulations that significant corrections to the redshift-space power spectrum in the Kaiser limit are expected at very large scales, $k \ga 0.01~h$Mpc$^{-1}$. In particular, we emphasize that the {\em shape} of the redshift-space power spectrum monopole {\em is not} a good approximation to the shape of the linear real-space power spectrum, even when  $k\le 0.1~h$Mpc$^{-1}$. We find that weakly non-linear effects tend to suppress monopole power increasingly with $k$, and more so for the quadrupole, supporting the argument discussed above that at least part of the transition from positive to negative quadrupole with increasing $k$ is due to large-scale effects, not just virial velocities. We briefly discuss the implications of these results for the determination of $\Omega_m$ and the reconstruction of the real-space power spectrum. 

Past work  along these lines was done by~\cite{1996MNRAS.279L...1F,1996MNRAS.282..767T,2000ApJ...537...12H}, who considered whether deviations from the Kaiser limit at large scales could be due to large-scale velocities. However, these relied on the  Zel'dovich approximation, which conserves momentum only to linear order, thus velocity fields are not described accurately enough to obtain reliable results (see e.g.~\cite{1998MNRAS.296...10H}).  Studies of the redshift-space power spectrum using dark matter numerical simulations have shown significant deviations at large scales from the Kaiser limit before (e.g.~\cite{1994MNRAS.267..785C,1998MNRAS.296...10H}), but these deviations have generally been blamed exclusively on virial velocities. In fact, as we discus here most of the large-scale velocity dispersion is due to weakly non-linear dynamics and thus has useful cosmological dependence on $\Omega_m$, $\sigma_8$ and the shape of the linear power spectrum that can be used to enhance constraints from galaxy clustering in redshift surveys.

This paper is organized as follows. In section~\ref{sec2} we derive the exact relation between real and redshift space two-point statistics, obtain the pairwise velocity PDF in the dispersion model, and discuss the recovery of the pairwise velocity PDF from clustering measurements. In section~\ref{Pzlin} we present the exact result for the redshift-space two-point correlation function in the case of Gaussian random fields and compare it to the Kaiser formula. We also present measurements of the pairwise velocity moments and discuss why Gaussianity is not a  good approximation even at large scales. In section~\ref{Pzlinls} we derive the large-scale limit of the redshift-space power spectrum and discuss how it differs from the standard approach in the literature. Section~\ref{theta1L} presents results from perturbation theory and N-body simulations on the weakly nonlinear evolution of velocity fields at large scales and why it differs substantially from that of the density field. Finally, in section~\ref{rsps} we present a simple model for the redshift-space power spectrum based on the results of previous sections and discuss the recovery of the real-space power spectrum. We summarize all the results in section~\ref{concl}. In paper II we present a calculation of the non-Gaussian terms in the evolution of pairwise velocities and their PDF.

\section{From Real Space to Redshift Space}
%
\label{sec2}

\subsection{Two-Point Statistics in Redshift Space and Pairwise Velocities}
\label{sec21}

In redshift-space, the observed radial position ${\bf s}$ of an object is given by its radial velocity, which reflects its true position due to the Hubble flow plus  ``distortions'' due to peculiar velocities. The mapping from its real-space position  ${\bf \x}$ is given by:

\beq
{\bf s}=\x - f \ \u(\x) {\hat z},
\label{zmap}
\eeq
where $f= {\rm d}\ln D/{\rm d}\ln a$ (with $D$ the growth factor and $a$ the scale factor) is a function of $\Omega_m$ alone for open models or flat models with a cosmological constant~\footnote{It is easy to show that in these cases $f$ obeys $(3/2)x=(a-bx) f + c x(x-1) f'+f^2$. For open models, $a=c=1$, $b=1/2$ and $f(\Omega_m) \approx \Omega_m^{3/5}$. For flat models with a cosmological constant, $a=2$, $b=3/2$, $c=3$ and $f(\Omega_m)  \approx \Omega_m^{5/9}$. We use the exact value from the ODE, which is straightforward to solve numerically.}, the scaled velocity field  ${\bf u}(\x) \equiv - \v(\x)/({\cal H} f)$, with $\v(\x)$ the peculiar velocity field, ${\cal H}^{-1}$ the comoving Hubble scale, and we have assumed the ``plane-parallel'' approximation, so that the line of sight is taken as a fixed direction, denoted by ${\hat z}$. The density field in redshift space is obtained by imposing mass conservation, i.e.

\beq
(1+\ds)\ d^3s=(1+\d)\ d^3x \label{mass},
\eeq
and thus we have in Fourier space, 

\beq
\dD(\k)+ \ds(\k) =  \int  \frac{d^3x}{(2\pi)^3} {\rm e}^{-i \k\cdot\x} \ 
{\rm e}^{i f k_z \u(\x)} \Big[1+ \d(\x) \Big]\label{d_s}.
\eeq

Note that this derivation is exact, it does not make any approximations about density or velocity fields; the only assumption is that we work in the plane parallel approximation, which is trivial to overcome by changing $k_z \u \rightarrow (\k \cdot \hat{x})\  (\ub \cdot \hat{x})$. Furthermore, since we are only using Eqs.~(\ref{mass}-\ref{d_s}), there is no reference to the Jacobian of the transformation from $\x$ to ${\bf s}$, Eq.~(\ref{d_s})  is valid even in regions where there is multistreaming. In other words, Eq.~(\ref{d_s}) is taking all mass elements at $\x$ and putting them at the corresponding ${\bf s}$, if different $\x$'s give rise to the same ${\bf s}$ they will be summed over as necessary. 

For the power spectrum, Eq.~(\ref{d_s}) gives:

\beqa
\dD(\k)+ P_s(\k) &=& \int \frac{d^3 r}{(2 \pi)^3} {\rm e}^{-i \k \cdot \r} \Big\langle
{\rm e}^{i f k_z \Delta \u }\  \nonumber \\  & & [1+ \d(\x) ] [1+ \d(\x') ] \Big\rangle
\label{Ps2},
\eeqa
where $\Delta \u \equiv \u(\x)-\u(\x')$ and $\r\equiv \x-\x'$. In configuration space we have

\beqa
1+ \xi_{s}(\spa,\spe)&=& \int d\rpa \ \Big\langle 
\d_{D}(\spa-\rpa+f\Delta\u)\nonumber \\ & &  [1+ \d(\x) ]\ [1+ \d(\x') ] \Big\rangle,
\label{xisfirst}
\eeqa
where the constraint given by the delta function takes a pair separated by line-of-sight distance $\rpa=(\x-\x')\cdot {\hat z}$ in real space to $ \spa $ in redshift-space as given by Eq.~(\ref{zmap}), with perpendicular separations unchanged, $\spe=\rpe$. Direct Fourier transformation of this equation yields Eq.~(\ref{Ps2}) for the power spectrum. We can write Eq.~(\ref{xisfirst}) in a form closer to that of the power spectrum by rewriting the delta function,

\beqa
1+ \xi_{s}(\spa,\spe)&=& \int \frac{d\rpa d\gamma}{2\pi} {\rm e}^{-i\gamma(\rpa-\spa)}\ \Big\langle {\rm e}^{if\gamma \Delta\u}\nonumber \\ & &  [1+ \d(\x) ]\ [1+ \d(\x') ] \Big\rangle,
\label{xis}
\eeqa

It is clear from Eqs.~(\ref{Ps2}) and (\ref{xis}) that the basic object of interest is the line-of-sight {\em pairwise velocity generating function}, $\M(\lambda,\r)$,

\beq
[1+\xi(r)]\ \M(\lambda,\r) \equiv \Big\langle {\rm e}^{\lambda\Delta\u}\  
 [1+\d(\x) ]\ [1+ \d(\x') ] \Big\rangle,
\eeq
where we are interested in $\lambda=i f k_z$ in Fourier space, or $\lambda=i f \gamma$ in configuration space. This generating function can be used to obtain the line-of-sight pairwise velocity moments, e.g. 

\beqa
v_{12}(\r) &\equiv& \Big( \frac{\partial \M}{\partial \lambda} \Big)_{\lambda=0} \label{v12} \\
\sigma^2_{12}(\r) &\equiv& \Big( \frac{\partial^2 \M}{\partial \lambda^2} \Big)_{\lambda=0}, \label{sig12}
\eeqa
give the mean and dispersion of the line-of-sight pairwise velocities~\footnote{Note that due to our normalization, velocities represented by $\u$ should be scaled by $-{\cal H} f$ to convert to km/s. This converts ``$u$-velocities" (which have units of Mpc~$h^{-1}$) to ``$v$-velocities" in km/s, see Eq.~(\ref{zmap}) and discussion below it. It will be clear from the context whether we are using ``$u$-velocities" (which are always accompanied by appropriate factors of $f$) or  ``$v$-velocities".}. The pairwise velocity probability distribution function (PDF), ${\cal P}(v)$, is obtained from the moment generating function by inverse Fourier transform~\footnote{The factor $f$ in the first argument of $\M$ in Eq.~(\ref{Pv}) takes into account the $\Omega_m$ dependence of peculiar velocities in linear theory. This is almost all there is, the remaining dependence is through a time average of $f^2/\Omega_m$, which is very weakly dependent on $\Omega_m$.},

\begin{equation}
\label{Pv}
{\cal P}(v,\r) = \int_{-\infty}^\infty \frac{d \gamma}{2\pi}\ {\rm e}^{-i \gamma v}\ 
\M(i \gamma f,\r).
\end{equation}
Notice that ${\cal P}(v)$ depends on scale through the scale-dependence of $\M$, and indeed $\int dv {\cal P}(v)\ v = f v_{12}(\r)$, $\int dv {\cal P}(v)\ v^2 = f^2 \sigma^2_{12}(\r)$, etc. From Eq.~(\ref{xis}) and~(\ref{Pv}) the redshift-space two-point correlation function can then be written as

\begin{equation}
\label{xiconv}
 1+\xi_{s}(\spa,\spe)= \int_{-\infty}^\infty  d\rpa\ [1+\xi(r)]\ {\cal P}(\rpa-\spa,\r),
\end{equation}
where $r^2 \equiv \rpa^2+\rpe^2$ and $\spe=\rpe$. The physical interpretation of this formula is clear: ${\cal P}$ maps the pairs at separation $\rpa$ to separation $\spa$ due to relative velocity $-{\cal H} (\rpa-\spa)$ [see Eq.~(\ref{zmap})] with probability ${\cal P}(\rpa-\spa,\r)$.  This type of relationship between the real and redshift space correlation functions is known as the streaming model~\cite{1980lssu.book.....P}, though it is commonly written in terms of $\xi$ rather than $1+\xi$. If ${\cal P}$ did not depend on scale, both formulations are equivalent, when there is scale dependence (as expected in any realistic scenario), the first term in the integral for ${\cal P}$ does not give unity, thus one should use  Eq.~(\ref{xiconv}) instead. In fact, this contribution to $\xi_s$ has a simple physical interpretation: it corresponds to redshift-space density fluctuations generated by velocity fluctuations in a uniform  (real-space) density, i.e. when $\xi=0$. If ${\cal P}$ did not depend on scale, random pairs are mapped into random pairs, scale dependence means that redshift-space correlations are created by taking random pairs in real space  and mapping them to redshift space differently at different scales.

The streaming model has been mostly  used at small non-linear scales by assuming ${\cal P}$ to be an exponential with zero streaming velocity and a {\em scale-independent isotropic} velocity dispersion~\cite{1983ApJ...267..465D}. At large scales, \cite{1995ApJ...448..494F}~showed that if one assumes the streaming model in phase space (with density and velocity fields coupled as in linear dynamics), it is possible to recover the Kaiser limit for the correlation function. We will stress in section~\ref{Pzlinls}, however, that the large-scale limit uses an {\em additional} assumption, that $\spa$ be much larger than the pairwise velocity dispersion. Fisher~\cite{1995ApJ...448..494F} also claims that in the linear regime the relationship between $\xi_{s}$ and $\xi$ can be reduced to the standard streaming model, i.e. as in Eq.~(\ref{xiconv}) with $1+\xi$'s replaced by $\xi$'s [see his Eq.~(26)]. This is incorrect, it suffices to see that if this were true all terms in $\xi_s$ would be proportional to $\xi$, in particular, such a result does not admit redshift distortions generated by correlated velocity fluctuations (where ${\cal P}$ depends on $\r$) in an unclustered distribution ($\xi=0$). 

The power spectrum and two-point correlation function in redshift space can be written in a similar form,

\beqa
P_{s}(\k) &=& \int \frac{d^3 r}{(2 \pi)^3} {\rm e}^{-i \k \cdot \r}
 \Big[ {\cal Z}(\lambda,\r)-1 \Big], 
\label{powerz} \\ & & \nonumber \\
\xi_{s}(\spa,\spe) &=& \int \frac{d\rpa d\gamma}{2 \pi} {\rm e}^{-i \gamma (\rpa-\spa)} \Big[ {\cal Z}(\lambda,\r)-1 \Big], \nonumber\\ &&\label{xiz}
\eeqa
where $\lambda=i f k_z,i f \gamma$ respectively and
\beq
{\cal Z}(\lambda,\r) \equiv [1+\xi(r)]\ \M(\lambda,\r).
\label{Zgen}
\eeq

It is important to note that the two-point correlation function is affected by redshift distortions for {\em all} configurations, even those perpendicular to the line of sight, since they are coming from different scales through the dependence of ${\cal P}$ on $\rpa$. It is however possible to project out redshift distortions by integrating along the line of sight,

\beqa
\xi_p(\rpe) &\equiv &\frac{2}{\rpe} \int_0^\infty  d\spa\ \xi_{s}(\spa,\rpe) \nonumber \\
&=& \frac{2}{\rpe} \int_0^\infty d\rpa\ \xi(\sqrt{\rpa^2+\rpe^2})\nonumber \\
& =& \pi \int P(k)\ \frac{J_0(k \rpe)}{k\rpe}\ d^3k,
\label{xip}
\eeqa

\noindent which sets $\gamma=0$ in Eq.~(\ref{xiz}).  This is only true in the plane-parallel approximation, where the concept of ``line of sight" is applicable. On the other hand,  the redshift-space power spectrum has the nice property, in the plane-parallel approximation, that transverse modes are unaffected by redshift distortions (a wave in the $\k_\perp$ direction is uniform in $z$ and thus unperturbed by the real-to-redshift space mapping), therefore $P_s(k_z=0,k_\perp)=P(k_\perp)$. 

\begin{figure}[t!]
\begin{center}
\includegraphics[width=0.5\textwidth]{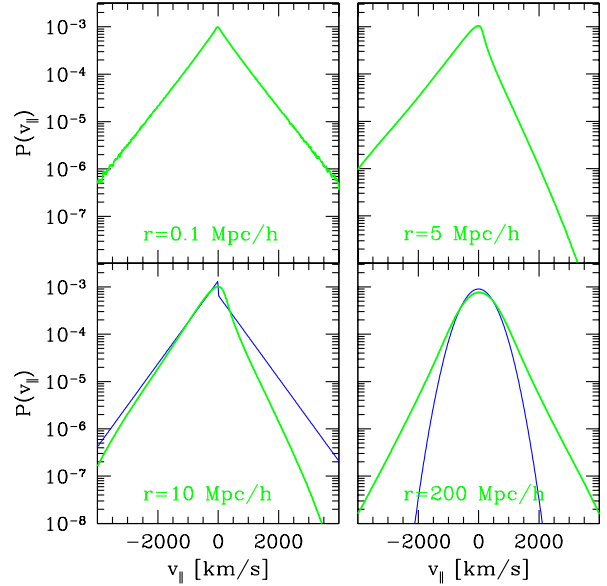}
\caption{The parallel to the line of sight pairwise velocity PDF at redshift $z=0$ for pairs separated by distance $r$, measured in the N-body simulations. In the bottom left panel, the discontinuous at the origin PDF (thin solid line) corresponds to that given by the dispersion model, Eq.~(\protect\ref{Pvdisp}) (ignoring the delta function at the origin). In the bottom right panel, the narrow distribution (thin solid line) corresponds to the prediction of linear dynamics, Eq.~(\protect\ref{Pv2}).}
\label{PDFpar}
\end{center}
\end{figure}

Figure~\ref{PDFpar} shows the pairwise velocity distribution ${\cal P}$ for pairs separated by distance $r$ along the line of sight, measured from the VLS simulation of the Virgo consortium~\cite{2001MNRAS.328..669Y}. This has $512^3$ dark matter particles in a 479 Mpc $h^{-1}$ box with a linear power spectrum corresponding to $\Omega_m=0.3$ (including $\Omega_b=0.04$ in baryons), $\Omega_\Lambda=0.7$, $h=0.7$ and $\sigma_8=0.9$. Due to the large number of pairs (in our measurements we use $32\times 10^{12}$ total pairs at scales between $0.1$ and $300$ Mpc$~h^{-1}$) and volume of the simulation, the statistical uncertainties are small enough that we do not plot error bars for clarity. On the other hand, one must keep in mind that neighboring points, separated by only 20 km/s, must be highly correlated.  

Note that at most scales $r\simeq 2-100$~Mpc$~h^{-1}$ the distribution is quite skewed (see also the central panel in Fig.~\ref{v12s12} below for a plot of the skewness $s_3$ as a function of scale). This arises as follows: the left tail ($v<0$) corresponds members of pairs approaching each other as they fall into an overdensity, the right tail ($v>0$) corresponds to members of pairs receding from each other as they empty underdense regions. Most pairs are not inside a void or falling coherently into a single structure, therefore the peak of ${\cal P}$ is close to $v=0$. The asymmetry between the left and right tail gives rise to a mean infall ($v_{12}<0$), that is, it is more probable to find ``coherent" pairs in overdense than underdense regions. 

Perhaps the most significant feature of ${\cal P}$ is that it has exponential wings {\em at all scales}, extending what was previously derived in the highly non-linear~\cite{1996MNRAS.279.1310S} and weakly non-linear~\cite{1998ApJ...504L...1J} regime, the prediction of linear perturbation theory (shown as the thin solid line in the bottom right panel) is {\em never} a good approximation, not even in the large-scale limit. The reasons for this are discussed in detail in section~\ref{fail}. The thin solid lines in the left bottom panel show the results of the dispersion model, although by assumption it has exponential tails, it is a poor match to simulations (even though $\sigma_p$ is fitted to the measured redshift-space power spectrum) and represents an unphysical (discontinuous and singular) distribution of pairwise velocities. See next section for details. 
  
\begin{figure}[t!]
\begin{center}
\includegraphics[width=0.5\textwidth]{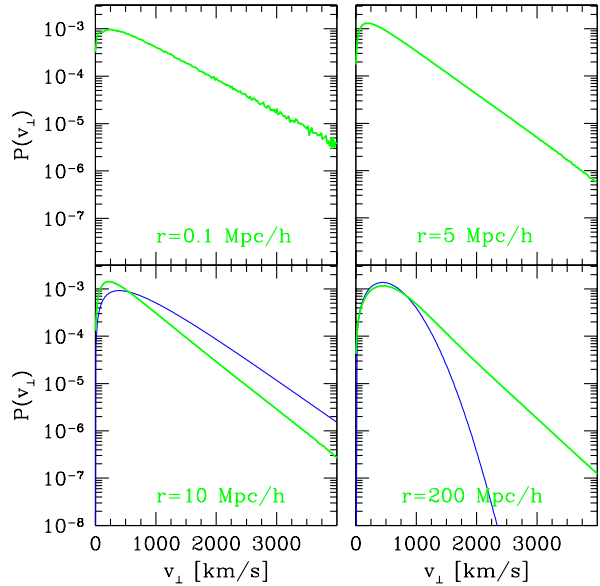}
\caption{Same as Fig.~\protect\ref{PDFpar} but for pairs perpendicular to the line of sight. 
The thin solid lines in the bottom panels are as in Fig.~\protect\ref{PDFpar}, the prediction of the dispersion model (left) and linear dynamics (right).}
\label{PDFperp}
\end{center}
\end{figure}

Figure~\ref{PDFperp} shows ${\cal P}$ for pairs separated by distance $r$ perpendicular the line of sight. In this case we define $v_{\perp}=\sqrt{v_x^2+v_y^2}$, then if $\widehat{\cal P}$ is the PDF for a perpendicular component of the velocity field (i.e. $v_x$ or $v_y$, it's the same by isotropy and even by symmetry) it follows that

\beq
{\cal P}(v_{\perp}) = 2\, v_{\perp} \int^{v_{\perp}}_{-v_{\perp}} \frac{dv_x }{\sqrt{v_{\perp}^2-v_x^2}} \ 
 \widehat{\cal P}(v_x)\ \widehat{\cal P}\Big(\sqrt{v_{\perp}^2-v_x^2}\Big).
 \label{Pperp}
 \eeq

\noindent For a Gaussian distribution $\widehat{\cal P}(v_x)=(2\pi \sigma^2)^{-1/2}\, {\rm e}^{-v_x^2/2\sigma_v^2}$ and thus ${\cal P}(v_{\perp})=(v_{\perp}/\sigma^2)\, {\rm e}^{-v_{\perp}^2/2\sigma^2}$. In this case ${\cal P}$ has zero skewness, by symmetry all odd moments vanish. Apart from this, the behavior of ${\cal P}$ is similar to the parallel case, the distribution is non-Gaussian at all scales and displays exponential tails. We now turn to a discussion of ${\cal P}$ in the dispersion model.

\subsection{The Dispersion Model}
\label{dismodel}

It is instructive to recast Eq.~(\ref{DMexp}) in terms of the full pairwise velocity distribution that it implies. There are two contributions to the pairwise PDF in this model, one given by the squashing factor, the other by the dispersion factor, with the total PDF being the convolution of both PDF's. The Fourier transform of the dispersion factor in Eq.~(\ref{DMexp}) corresponds to a pairwise velocity PDF that is exponential, that is

\beq
{\cal P}_{\rm disp}(v)= \frac{{\rm e}^{-|v|\sqrt{2}/\sigma_p}}{\sqrt{2}\sigma_p},
\label{Pexpo}
\eeq

The squashing factor in the Kaiser limit corresponds to a delta function PDF, see Eq.~(\ref{PvK}) below for a derivation. Performing the convolution of this with Eq.~(\ref{Pexpo}) leads to the pairwise PDF in the dispersion model,

\beqa
{\cal P}(v) =  \frac{{\rm e}^{-|v|\sqrt{2}/\sigma_p}}{\sqrt{2}\sigma_p} &\Big[& 1 \mp \frac{\sqrt{2}}{\sigma_p} fv_{12} + \frac{2 f^2 \psi_v}{\sigma_p^2} \nonumber \\ &&\times [1-\sqrt{2}\sigma_p\ \delta_D(v)] \Big],
\label{Pvdisp}
\eeqa

where the $+$ sign corresponds to $v>0$, and the $-$ sign to $v<0$ (opposite for ``$v$-velocities" shown in Fig.~\ref{PDFpar}), and $\psi_v$ denotes the velocity-velocity correlation function in linear dynamics, $\psi_v=\psi_{\perp}+\nu \Delta\psi$, see Eqs.~(\ref{corrvv}) and~(\ref{Dpsi}) for explicit expressions. Note that the resulting PDF is singular at the origin, and in addition has a jump discontinuity at $v=0$ which is proportional to $v_{12}$. The bottom left panel in Figs.~\ref{PDFpar}  and~\ref{PDFperp} illustrate this result (omitting the singular term at $v=0$) and compares it to the measurements in numerical simulations for a separation of $r=10$~Mpc/$h$. We have fitted the value of $\sigma_p$, as it is normally done, to the measured quadrupole to monopole ratio of the redshift-space power spectrum. Despite this fit to the power spectrum, the resulting PDF does not fit the simulation results. This is hardly surprising, since the dispersion model Eq.~(\ref{DMexp}) makes unphysical predictions for the pairwise velocity PDF, see Eq.~(\ref{Pvdisp}).

\subsection{Recovery of the Pairwise Velocity PDF from Redshift-Space Two-Point Statistics}
\label{PkToPDF}

Given the relationship between the redshift-space and real-space correlation function through the pairwise velocity PDF, Eq.~(\ref{xiconv}), it is natural to ask whether one can recover information about the PDF from clustering measurements. The problem is that there is no single PDF involved in Eq.~(\ref{xiconv}), but rather an infinite number of PDF's corresponding to different scales and angles of the velocities with respect to the line joining the pair. If there was no scale dependence and anisotropy, all the PDF's are the same and Eq.~(\ref{xiconv}) becomes a convolution, thus one can find the PDF by deconvolution. In other words, due to the scale dependence of the pairwise velocity PDF, Eq.~(\ref{xiconv}) is not really a convolution; this implies that the redshift-space power spectrum for modes parallel to the line of sight {\em is not} the real-space power spectrum multiplied by the generating function ${\cal M}$. Instead,  from Eq.~(\ref{powerz}) we get 

\beqa
P_s(\k) &=& P(k)+ \widetilde{\cal M}(ifk_z,\k) \nonumber \\ &+&  
\int d^3q\ \widetilde{\cal M}(ifk_z,\k-\q)\ P(q),
\label{convconv}
\eeqa
where $\widetilde{\cal M}$ [recall that ${\cal M}(ifk_z,\r) = \int {\cal P}(v,\r) {\rm e}^{ivk_z} dv$] is basically the double-Fourier transform of ${\cal P}(v,\r)$,

\beqa
\widetilde{\cal M}(\lambda,\p) &\equiv&  \int \frac{d^3 r}{(2 \pi)^3} {\rm e}^{-i \p \cdot \r}\ 
[{\cal M}(\lambda,\r)-1] \nonumber \\ &=&
\int  \frac{ \Big\langle ({\rm  e}^{\lambda \Delta\u}-1) (1+\d)(1+\d') \Big\rangle}{1+\xi}, \nonumber \\ &&
 \times \ {\rm e}^{-i \p \cdot \r}\  \frac{d^3 r}{(2 \pi)^3}  
\eeqa

\noindent except that we substract the zero mode ${\cal M}(0,\r)=1$, thus 
$\widetilde{\cal M}(0,\p)=0$. For example, in the Kaiser limit we have 
\beq
\widetilde{\cal M}(ifk_z,\p) \rightarrow 2f \frac{k_zp_z}{p^2}\ P_{\delta\theta}(p) + f^2 \frac{k_z^2p_z^2}{p^4}\ P_{\theta\theta}(p),
\label{DMk}
\eeq
where $P_{\delta\theta}$ denotes the density-velocity divergence power spectrum and $P_{\theta\theta}$ is the velocity divergence power spectrum. In linear PT, $P=P_{\d\d}=P_{\d\te}=P_{\te\te}$, but we will keep the distinction because weakly non-linear corrections are significant at large scales, see section~\ref{theta1L}.

If we assume that all pairwise moments have no scale dependence and are isotropic, which implies that odd moments vanish (since they must be anisotropic, by symmetry odd moments vanish when $\r \cdot \hat{z}=0$), $\widetilde{\cal M}(ifk_z,\p) = [{\cal M}(ifk_z)-1] \delta_D(\p)$ and thus  $P_s(k\hat{z})=P(k)\ {\cal M}(ifk)$. Note that in this case ${\cal M}(ifk)$ is real  because odd moments vanish, however  in general ${\cal M}(ifk)$ is complex. By taking (even number of) derivatives with respect to $\lambda$ of  $P_s(\sqrt{-\lambda^2}\hat{z})/P(\sqrt{-\lambda^2})={\cal M}(\lambda)$ one can generate all (even) moments and thus find the (symmetric by assumption) PDF by inverse Fourier transform.

Galaxy redshift surveys show that $P_s(k\hat{z})/P(k)$ is very close to a Lorentzian, and this has been interpreted as evidence for an exponential pairwise velocity PDF~\cite{1998ApJ...494L.133L,2002ApJ...567L...1L}. However, realistically one cannot neglect anisotropy, since we know that odd moments must be non-zero, in particular there are infall velocities ($v_{12}\neq 0$) and skewness. The infall velocities are small compared to the dispersion at small, non-linear scales, however the skewness is expected to  be significant except in the highly non-linear regime, see Fig.~\ref{v12s12} below~\cite{1994ApJ...431..559Z,1994MNRAS.267..927F,1998ApJ...504L...1J,2001MNRAS.322..901S}. By construction, since power spectra are functions of only $k^2$ and $k_z^2$ by statistical isotropy, using $P_s(k\hat{z})/P(k)$ as a generating function (with $k=-i\lambda$) only generates even moments and thus a {\em symmetric} answer for its PDF, even if the actual PDF is asymmetric~\footnote{In fact, it is apparent  from plots in~\cite{1998ApJ...494L.133L,2002ApJ...567L...1L} that the answer returned by this method for the PDF is manifestly symmetric, e.g. all the same noisy features appear at $v$ and $-v$.}. 

To see what this method actually recovers we go back to Eq.~(\ref{convconv}) and write explicitly that the redshift-space power spectrum depends on the magnitude of $k$ and $k_z$,

\beqa
P_s(k,k_z) &=&P(k)+ \frac{1}{2}[\widetilde{\cal M}(ifk_z,\k)+\widetilde{\cal M}(-ifk_z,-\k)] \nonumber \\ 
&+&  
\int d^3q\ \frac{1}{2}[\widetilde{\cal M}(ifk_z,\k-\q)\nonumber \\ 
&+&\widetilde{\cal M}(-ifk_z,-\k-\q)]\ P(q), 
\label{convconv2}
\eeqa
that is, we average together waves with opposite wavevectors. In this method the moment generating function is identified with the ratio $P_s(k\hat{z})/P(k)$ (where we must replace $k$ by $-i\lambda$ in our convention), thus it becomes

\beqa
{\cal G}(\lambda) &\equiv& 1+
\int \frac{d^3r}{(2\pi)^3}\ [1+\xi(r)] \nonumber \\ &\times & 
\frac{[{\cal M}(\lambda,\r)-1]{\rm e}^{-\lambda z}+
[{\cal M}(-\lambda,\r)-1]{\rm e}^{\lambda z}}{2P(\sqrt{-\lambda^2})}.
\nonumber \\ & & 
\label{genw}
\eeqa

It is easy to check that if ${\cal M}(\lambda)$ does not depend on $\r$, 
${\cal G}(\lambda)=M(\lambda)$ and thus one can recover by inverse Fourier transform of 
${\cal G}(\lambda)$ the PDF of pairwise velocities, as discussed above. However, in the realistic case with scale dependence and anisotropy, the inverse Fourier transform of 
${\cal G}(\lambda)$ {\em does not} give the actual PDF. To illustrate  this, let us calculate the first two non-vanishing moments (second and fourth) of this symmetric ``pseudo-PDF",

\beq
\Big( \frac{\partial^2 {\cal G}}{\partial \lambda^2}\Big)_{\lambda=0} = 
\int \frac{d^3r}{V\bar{\xi}}\ [\sigma_{12}^2(\r)-2z\ v_{12}(\r)]\ [1+\xi(r)],
\label{m2w}
\eeq

\beqa
\Big( \frac{\partial^4 {\cal G}}{\partial \lambda^4}\Big)_{\lambda=0} &=& 
\int \frac{d^3r}{V\bar{\xi}}\ [m_{12}^{(4)}(\r)-4z\ m_{12}^{(3)}(\r) \nonumber \\ 
&+&6z^2\ \sigma_{12}^2(\r)-4z^3\ v_{12}(\r)]\ [1+\xi(r)],\nonumber \\  & &
\label{m4w}
\eeqa

where $\bar{\xi} \equiv V^{-1}  \int d^3r \xi(r)$ is the average of the correlation function, and 
the integrals over the volume $V$ are cutoff at some large scale (depending on the size of the survey and the practical implementation of the method). In  Eqs.~(\ref{m2w}-\ref{m4w}) $m_{12}^{(3)}$ denotes the third moment of the actual PDF and $m_{12}^{(4)}$ its fourth moment. It is clear from these equations that the moments of this pseudo-PDF are weighted versions of combinations of {\em several} moments of the true PDF, so their value is not straightforward to interpret. From Eq.~(\ref{m2w}) we see, for example, that the effective value of the velocity dispersion $\sigma^2_{\rm eff}$ picked up by this method is given by 

\beq
\sigma^2_{\rm eff} = \int \frac{d^3r}{V\bar{\xi}}\ [1+\xi(r)] \ \Big[\frac{2}{3} \sigma_{\perp}^2(r)+\frac{1}{3}\sigma_\parallel^2(r)-\frac{2}{3}r\ \tilde{v}_{12}(r)\Big],
\label{sigeff}
\eeq

where we used that by symmetry $\sigma^2_{12}(\r) = \sigma_{\perp}^2(r) + \nu^2 [\sigma_\parallel^2(r)-\sigma_{\perp}^2(r)]$ and $v_{12}(\r)=\nu\ \hat{v}_{12}(r)$, with $\nu = z/r$. Therefore, we see that the effective value of the velocity dispersion is a weighted version  of the underlying velocity dispersion {\em minus} a contribution due to the mean streaming  (recall in our convention here $\tilde{v}_{12}(r)>0$, these are ``$u$-velocities"), therefore one expects this method to yield a biased low value of the weighted [by $r^2(1+\xi)$] mean of the dispersion if not corrected for infall, as stressed by~\cite{1998ApJ...503..502J} and more recently in~\cite{2003MNRAS.346...78H}. Note however that~\cite{2003MNRAS.346...78H} also interpret the pseudo PDF as the actual PDF of pairwise velocities, and they do not include skewness into their treatment.

\subsection{The pairwise velocity PDF in terms of its building blocks}
\label{CumExp}

We now discuss how to use cumulant expansions to evaluate the pairwise moment generating function or ${\cal Z}(\lambda,\r) $ in Eq.~(\ref{Zgen}) in terms of its building blocks, the cumulants. The starting point is the property between the moment ${\cal M}(\,{\bf j})$ and cumulant ${\cal C}(\,{\bf j})$ generating functions for a set of fields which we group together in a vector field ${\bf A}$,

\beq
{\cal M}(\,{\bf j})=\big\langle {\rm e}^{\,{\bf j} \cdot{\bf A}}\big\rangle = \exp \big\langle {\rm e}^{\,{\bf j} \cdot{\bf A}}\big\rangle_c = \exp {\cal C}(\,{\bf j}),
\eeq

where ${\bf A}=\{A_1, \ldots, A_n\}$ and similarly for ${\bf j}$. Derivatives of ${\cal C}$ generate all the connected correlation functions. By taking derivatives with respect to appropriate components of the vector ${\bf j}$, it follows in particular that

\beqa
\big\langle {\rm e}^{j_1  A_1} A_2 \big\rangle &=& \big\langle {\rm e}^{j_1  A_1} A_2 \big\rangle_c \exp \big\langle {\rm e}^{j_1  A_1} \big\rangle_c, \\
\big\langle {\rm e}^{j_1  A_1} A_2 A_3 \big\rangle &=& 
\Big[ \big\langle {\rm e}^{j_1  A_1} A_2 A_3 \big\rangle_c 
+ \big\langle {\rm e}^{j_1  A_1} A_2\big\rangle_c \nonumber \\ & & 
\big\langle {\rm e}^{j_1  A_1} A_3 \big\rangle_c 
\Big] \exp \big\langle {\rm e}^{j_1  A_1} \big\rangle_c.
\eeqa

Using $j_1=\lambda$, $A_1=\Delta\u$, $A_2=\d(\x)\equiv \d$ and $A_3=\d(\x')\equiv \d'$, 
this leads to the {\em exact} expression,

\beqa
{\cal Z}(\lambda,\r) &=&\ \exp \big\langle {\rm e}^{\lambda  \Delta\u} \big\rangle_c
  \Big[ 1+ \Big\langle {\rm e}^{\lambda  \Delta\u} \d \Big\rangle_c  
 \nonumber \\ &+& \Big\langle {\rm e}^{\lambda  \Delta\u} \d' \Big\rangle_c 
 + \Big\langle {\rm e}^{\lambda  \Delta\u} \d \Big\rangle_c  
  \Big\langle {\rm e}^{\lambda  \Delta\u} \d' \Big\rangle_c \nonumber \\ &
  +&  
\Big\langle {\rm e}^{\lambda  \Delta\u} \d \d' \Big\rangle_c \Big] 
\label{Zgen2}
\eeqa

Note that the overall factor in this expression is the moment generating function for the line of sight velocity differences, and it is a {\em volume weighted} quantity (as opposed to the pairwise velocity PDF which is mass weighted, by densities at $\x$ and $\x'$). This velocity-difference PDF is not sensitive to galaxy biasing, since it does not depend on the density field and even if there is velocity bias inside dark matter halos this is a small effect~\cite{1990ApJ...352L..29C,2000ApJ...539..561C,2003ApJ...593....1B}  and halos are in addition suppressed by volume weighting due to their small size. Therefore, the velocity-difference PDF depends on weakly nonlinear dynamics and thus can be modeled (almost) exclusively in terms of cosmological parameters.

It is straightforward to evaluate Eq.~(\ref{Zgen2}) in the linear regime, for Gaussian fluctuations. In this case, the velocity is proportional to the density, whose only non-zero cumulant is the second and thus 

\beqa
{\cal Z}_{\rm G}(\lambda,\r)& =&  \Big[ 1+ \xi(r)  + \lambda \Big\langle  \Delta\u [\d + \d' ]\Big\rangle+ \lambda^2 \Big\langle  \Delta\u \d \Big\rangle \nonumber \\ 
& &\Big\langle  \Delta\u \d' \Big\rangle  \Big] \times \ \exp \frac{\lambda^2}{2} \big\langle \Delta u_z^2 \big\rangle \label{Zlin}
\eeqa

Notice that even in this case, the resulting expression is non-linear in the amplitude of correlation functions and does not involve terms of the same order in linear perturbation theory. Even though fluctuations are assumed to obey the linear dynamics, the non-linear nature of the redshift-space mapping leads to a somewhat more complicated picture. We will explicitly evaluate Eq.~(\ref{Zlin}) in the next section. Non-linear effects due to dynamics lead to significant deviations from the predictions of Eq.~(\ref{Zlin}), even at large scales, we discuss this below. An evaluation of  Eq.~(\ref{Zgen2}) is given in paper II.

\section{The Redshift-Space Power Spectrum and Correlation Function in Linear Dynamics}
\label{Pzlin}
\subsection{Pairwise velocity moments}
\label{PmomG}

We now give an explicit evaluation of Eq.~(\ref{Zlin}). Using symmetry considerations, the velocity correlation function can be written as 

\beqa
\lexp u_i(\x+\r/2)\ u_j(\x-\r/2) \rexp &=& \psi_{\perp}(r)\ \delta_{ij} \nonumber \\ &+&
[\psi_{\parallel}(r)-\psi_{\perp}(r)]\ \frac{r_{i}r_{j}}{r^{2}},\nonumber \\ & &
\label{corrvv}
\eeqa

where $\psi_{\parallel}(r)$ and $\psi_{\perp}(r)$ are the velocity correlation functions parallel and perpendicular to the line of sight, respectively. They are related to the velocity divergence power spectrum $P_{\theta\theta}(k)$ through~\cite{1988ApJ...332L...7G}

\beq
\psi_{\perp}(r) = \int \frac{P_{\theta\theta}(k)}{k^{2}}\ \frac{j_{1}(kr)}{kr}\ d^{3}k,
\eeq

\beq
\psi_{\parallel}(r) = \int \frac{P_{\theta\theta}(k)}{k^{2}}\ [j_{0}(kr)-
2 \frac{j_{1}(kr)}{kr}]\ d^{3}k,
\eeq

\noindent where $j_\ell(x)$ is the usual spherical Bessel function, and we assumed a potential flow, which implies $\psi_{\parallel}(r)=d(r \psi_{\perp})/dr)$ and $\psi_{\perp}(r) \geq \psi_{\parallel}(r)$. The variance of velocity differences reads,

\beq
\big\langle \Delta u_z^2 \big\rangle=  2\Big( \sigma_{v}^{2} -
\psi_{\perp}(r) +\frac{z^2}{r^2}\ \Delta \psi(r)  \Big),
\label{DuzSQ}
\eeq

\noindent which leads to the volume-weighted velocity difference moment generating function

\beqa
{\cal Z}_{0}(\lambda,\r)  &\equiv& \exp \frac{\lambda^2}{2} \big\langle \Delta u_z^2 \big\rangle
\nonumber \\ &=&  \exp  \lambda^{2}\Big( \sigma_{v}^{2} -
\psi_{\perp}(r) +\frac{z^2}{r^2}\ \Delta \psi(r) \Big), \nonumber \\ & & 
\eeqa

\noindent where 
\beq
\Delta \psi(r) \equiv \psi_{\perp}(r)-\psi_{\parallel}(r)= 
\int \frac{P_{\theta\theta}(k)}{k^{2}}\ j_{2}(kr)\ d^{3}k,
\label{Dpsi}
\eeq
 
\noindent and the one-dimensional linear velocity dispersion $\sigma_{v}^{2}$ is given by

\beq
\sigma_{v}^{2} \equiv \frac{1}{3} \int \frac{P_{\theta\theta}(k)}{k^{2}}\ d^{3}k.
\eeq 

Note that as $r \rightarrow 0$, $\psi_{\parallel}(r) = \psi_{\perp}(r)=\sigma_{v}^{2}$, and then ${\cal Z}_{0}(\lambda,0) =1$, as expected from its definition. On the other hand, as $r \rightarrow \infty$, $\psi_{\parallel}(r)$, $\psi_{\perp}(r)   \rightarrow 0$, and then 
${\cal Z}_{0}(\lambda,\infty ) \approx \exp -\lambda^{2} \sigma_{v}^{2}$. To evaluate the prefactors in Eq.~(\ref{Zlin}), we use that [see Eq.~(\ref{v12})]

\beqa
\Big\langle  \Delta\u [\d(\x) + \d(\x') ]\Big\rangle &=& v_{12}(\r)\ [1+\xi(r)] \nonumber \\ &
 =& 2 \frac{z}{r} \int \frac{P_{\d\theta}(k)}{k}\ j_{1}(kr)\ d^{3}k, \nonumber \\ & &
\eeqa
and
\beq
\Big\langle  \Delta\u \d(\x) \Big\rangle = \Big\langle  \Delta\u \d(\x') \Big\rangle = 
\frac{1}{2}\ v_{12}(\r)\ [1+\xi(r)],
\eeq
then
\beqa
{\cal Z}_{\rm G}&=& [1+\xi(r)]\ \Big[ 1+ \lambda v_{12}(\r) + \frac{\lambda^2}{4} v_{12}^2(\r) [1+\xi(r)] \Big]\nonumber \\ & \times & \exp \lambda^{2}\Big( 
\sigma_{v}^{2} -
\psi_{\perp}(r) +\frac{z^2}{r^2}\ \Delta \psi(r)  \Big). \label{Zlinear}
\eeqa

\subsection{The Failure of Gaussianity}
\label{fail}

It is important to note that, although the large-scale limit of $v_{12}$ is well described by linear dynamics (see e.g. \cite{1999ApJ...518L..25J,2001MNRAS.325.1288S}) the same is {\em not} true for the pairwise dispersion, indeed we have [$\d\equiv\d(\x)$, $\d'\equiv\d(\x')$]

\beqa
\sigma_{12}^2 \, (1+\xi)&=& \Big\langle  \Delta\u^2 (1+\d)(1+\d')\Big\rangle \nonumber \\ & =& \langle  \Delta\u^2 \rangle \, (1+\xi) + \langle  \Delta\u^2 (\d+\d')\rangle \nonumber \\ & + & 
\langle  \Delta\u^2  \d \d' \rangle_c.
\label{sig12exp}
\eeqa

In linear dynamics, Gaussianity implies that the last two terms vanish; however, in reality the third moment term contributes a constant in the large-scale limit ($r=|\x-\x'|\rightarrow\infty$) that adds in quadrature to the contribution of the first term (we evaluate this term in paper II). Therefore, {\em linear theory never gives a good approximation to the second moment of pairwise velocities}. That there are non-Gaussian corrections should be of no surprise since pair weighting means the second moment of pairwise velocities involves up to fourth moments~\cite{1998ApJ...504L...1J}, the interesting aspect here is that even in the large-scale limit non-Gaussian terms persist, e.g. $ \langle  \u^2 \d\rangle$ contributes a constant at large scales.

\begin{figure}[t!]
\begin{center}
\includegraphics[width=0.5\textwidth]{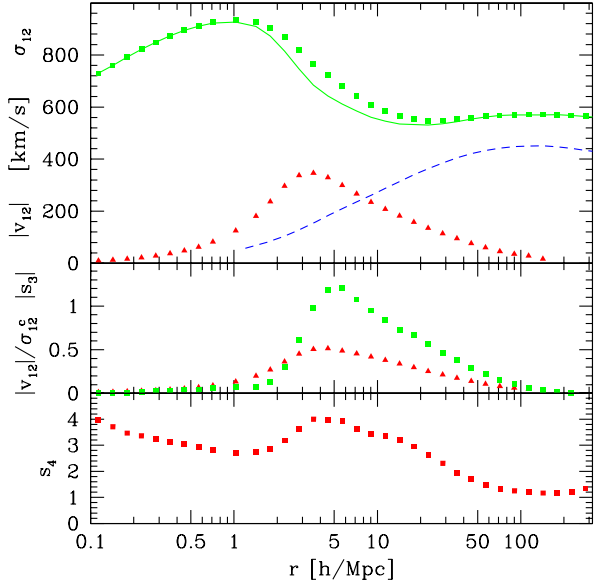}
\caption{Moments of pairwise velocities parallel to the line conecting the pair as a function of scale. {\em Top panel:} pairwise dispersion $\sigma_{12}$ (squares) as a function of scale, its connected piece $\sigma_{12}^{\, c}$ (solid line), and the mean infall $v_{12}$ (triangles). The dashed line denotes the predicted $\sigma_{12}$ in linear dynamics. {\em Middle panel:} dimensionless measure of infall ($|v_{12}|/\sigma_{12}^{\, c}$, triangles) compared to the skewness of the pairwise velocity PDF (squares); the skewness dominates at most scales. {\em Bottom panel:} kurtosis as a function scale, note that it does not vanish at large scales and $s_4>1$ at all scales; the pairwise velocity PDF is strongly non-Gaussian at all scales, see Fig.~\protect\ref{PDFpar}. For reference, an exponential distribution has $s_3=0$ and $s_4=3$.}
\label{v12s12}
\end{center}
\end{figure}

The top panel in Fig.~\ref{v12s12} illustrates this point, where $\sigma_{12}$ is shown as a function of scale for the N-body measurements (square symbols) and linear theory (dashed lines). All quantities in this figure refer to velocity components parallel to the separation vector of the pair. It is also important to note that the dependence of $\sigma_{12}$ on scale is {\em opposite} in the linear case (decreasing at smaller scales) than in the simulations (though at scales $r\la 1$ Mpc/h,  $\sigma_{12}$ starts decreasing in the N-body results). This is also a feature of the Gaussian restriction of linear dynamics, as we shall discuss in paper II, and it implies that the dispersion effect on the two-point correlation function or power spectrum will be significantly underestimated. Physically, in the Gaussian case as $r$ is decreased the velocity field is more correlated and thus $\langle  \Delta\u^2 \rangle$ decreases; since no correlations between density and velocity squared are incorporated in linear theory, it is impossible to see that the velocity of pairs in regions of larger overdensity are fluctuating more; this is described by the non-Gaussian third and fourth terms in Eq.~(\ref{sig12exp}). 

The other panels in Fig.~\ref{v12s12} show how important non-Gaussianity of the pairwise PDF is. The central panel compares the skewness $s_3$ to the dimensionless measure of infall, $v_{12}/\sigma_{12}^c$, where $\sigma_{12}^c$ is the connected second moment shown by solid lines in the top panel. This shows that the skewness is more important than infall at most scales (and by a large factor at scales where infall is most important). Therefore, modeling the pairwise PDF with infall but no skewness, as it is often done (see~\cite{1995ApJ...448..494F} for an exception), is not a good approximation. Finally, the bottom panel shows the kurtosis $s_4$ as a function of scale, this quantifies that the pairwise PDF is strongly non-Gaussian ($s_4>1$) at all scales, and it is basically a manifestation of the exponential wings seen in Figs.~\ref{PDFpar} and~\ref{PDFperp} at all separations. 

Why is the Gaussian limit of the pairwise velocity PDF never reached at large separations? The reason is that the relevant quantity is the (density-weighted) {\em difference} in velocities. At a given separation $r$ the velocity difference does not receive contributions from modes with wavelengths much larger than $r$, since those give the same velocity to $\x$ and $\x'$. For wavelengths smaller than $r$ the contribution of modes is down-weighted only by $k^{-1}$ (independent of $r$ in the $r\rightarrow\infty$ limit); therefore even at large separations one is sensitive to nonlinearities. In other words, at large separations the velocities are uncorrelated and thus the pairwise velocity generating function factorizes into individual particle velocity generating functions. These are sensitive to non-linearities, i.e. there is no ``large scale" in that problem. Thinking in terms of the halo model, at large scales the pairwise dispersion is due to particles in different halos, each of which has its own (independent) one-point dispersion due to virial (``non-linear") and halo (``linear") motions, these contributions will add in quadrature to give the full dispersion (see~\cite{2001MNRAS.325.1288S}). We caution, however, that this split is not straightforward, halo motions are not well described by linear theory (their pairwise PDF in the large-scale limit is not exactly Gaussian, see~\cite{2003MNRAS.343.1312H}).

In \cite{1998ApJ...504L...1J} it is argued that exponential tails in the pairwise PDF are generated by pair weighting; although this is in part important, it is not the whole story. We show in paper II that the velocity difference PDF (which is {\em  volume} weighted) also has exponential tails in the large-scale limit, for the reasons discussed above. Of course, pair weighting helps build non-Gaussianity and it is responsible for the deviations in $\sigma_{12}$ from linear theory at large scales.

\subsection{The Exact Result for Gaussian Random Fields}
\label{Pzlingen}

Even though Gaussianity is not a good approximation to describe the statistics of pairwise velocities, it is instructive to discuss the redshift-space correlation function in the Gaussian case, both as a starting point for more accurate calculations and to discuss the regime of validity of the Kaiser limit. 

The only assumptions in deriving Eq.~(\ref{Zlinear}) are that fluctuations are Gaussian and velocity flows are potential, i.e. there is no assumption about the amplitude of fluctuations (in practice, of course, Gaussianity follows only if fluctuations are vanishingly small). It is easy to write down explicitly the pairwise velocity PDF obtained from using Eq.~(\ref{Zlinear}) in Eq.~(\ref{Pv}),

\begin{widetext}
\begin{equation}
\label{Pv2}
{\cal P}(v) = \frac{1}{\sqrt{2\pi}f\big\langle \Delta u_z^2 \big\rangle^\frac{1}{2}}\ \exp \Bigg(\frac{-v^2}{2f^2\big\langle \Delta u_z^2 \big\rangle} \Bigg) \ \Bigg[ 1+ \frac{v\ v_{12}}{f \big\langle \Delta u_z^2 \big\rangle} + \frac{1}{4} \Big(\frac{v^2}{f^2\big\langle \Delta u_z^2 \big\rangle}-1\Big)
\ \frac{v_{12}^2}{\big\langle \Delta u_z^2 \big\rangle}\ (1+\xi) \Bigg].
\end{equation}
\end{widetext}

\noindent This is {\em not} a Gaussian distribution (except when $v_{12}=0$ at large or small scales, or at all scales for separations perpendicular to the line of sight), although close to its peak it is well approximated by a Gaussian centered at $v=fv_{12}$. Note however that the velocity difference PDF {\em is} Gaussian, being the prefactor outside the square brackets. The second and higher cumulants of the pairwise velocity PDF are e.g. 

\beqa
\langle v^2 \rangle_c &=& \sigma_u^2+\frac{\xi-1}{2}\, \langle v \rangle^2, \ \ \ \ \ 
\langle v^3 \rangle_c = \frac{1-3\xi}{2}\, \langle v \rangle^3, \nonumber \\
\langle v^4 \rangle_c &=& -\frac{3}{4}(1-6\xi+\xi^2)\, \langle v \rangle^4, \ \ \ \ \ 
\eeqa

where $\langle v \rangle = f v_{12}$, and $\sigma_u^2 \equiv f^2 \langle  \Delta u_z^2 \rangle $ is the variance of the distribution of velocity differences. The non-Gaussianity is induced solely by the non-linearities in the mapping from real to redshift space. The two-point function can be written using Eq.~(\ref{xiconv}),

\beqa
1+ \xi_{s}(\spa,\spe)& =& \int_{-\infty}^\infty \frac{d\rpa\ \ 
{\rm e}^{-\frac{1}{2}x^2}}{\sqrt{2\pi}f\ \big\langle \Delta u_z^2 \big\rangle^\frac{1}{2}}\ \ 
[1+\xi(r)]\nonumber \\
&\times&\Big[1+x\ u_{12} + \frac{(x^2-1)}{4}\ u_{12}^2\ (1+\xi) \Big], \nonumber \\& & 
\label{xisexlin}
\eeqa

\noindent where
\beq
x \equiv \frac{\rpa-\spa}{f \big\langle \Delta u_z^2 \big\rangle^\frac{1}{2}}, 
\ \ \ \ \ \ \ \ 
u_{12} \equiv \frac{v_{12}}{\big\langle \Delta u_z^2 \big\rangle^\frac{1}{2}}, 
\ \ \ \ \ \ \ \
r^2 \equiv \rpa^2 +\spe^2
\eeq

\noindent This is the exact result in the Gaussian limit, and has been obtained before by Fisher (\cite{1995ApJ...448..494F}, Eq.~20) by integrating the four dimensional joint Gaussian PDF for $\d(\x)$, $\d(\x')$, $\u(\x)$ and $\u(\x')$. See also~\cite{2003SPIE.4834..173L,2001MNRAS.327..577B}. The method described in section~\ref{CumExp} is an alternative way of obtaining the same result with considerably less algebra, and the advantage that also holds in the non-Gaussian case  provided the correlators in Eq.~(\ref{Zgen2}) can be calculated. The analogous result for the power spectrum is ($\nu \equiv z/r$, $k_\perp\equiv k \sqrt{1-\mu^2}$, $r_\perp\equiv r \sqrt{1-\nu^2}$), 

\begin{widetext}
\beq
\label{PksG }
P_s(k,\mu) = \frac{1}{2\pi^2} \int_0^\infty r^2 dr \int_0^1 d\nu\ 
J_0\big(k_\perp r_\perp \big)\ \Big[  \cos (kr\mu\nu)\ \ 
[{\cal Z}_{\rm G}^{\rm even}(\lambda,r,\nu)-1]\ +
 \sin (kr\mu\nu)\ \ {\cal Z}_{\rm G}^{\rm odd}(\lambda,r,\nu) \Big], 
\eeq
\end{widetext}

\noindent which involves a 2D rather than 1D integration. Here ${\cal Z}_{\rm G}^{\rm odd}(\lambda,r,\nu)$ corresponds to the term proportional to $\lambda$ in Eq.~(\ref{Zlinear}), and ${\cal Z}_{\rm G}^{\rm even}(\lambda,r,\nu)$ is the rest. In order to obtain power spectrum multipoles it is sometimes more convenient to calculate first multipoles of the correlation function, 

\begin{equation}
\label{ xiell}
\xi_\ell(r) = \frac{(2\ell+1)}{2} \int_{-1}^1 d\nu \ \xi_s(r,\nu)\ \pl_\ell(\nu),
\end{equation}
where $\pl_\ell$ denote the Legendre polynomials,  and then using the plane-wave expansion ($\mu \equiv k_z/k$, $\nu \equiv z/r$)

\beq
{\rm e}^{-i \k \cdot \r} = \sum_{\ell =0}^\infty (-i)^\ell (2\ell+1) \ 
j_\ell(kr)\  \pl_\ell(\mu)\  \pl_\ell(\nu),
\eeq
obtain from them the power spectrum multipoles,

\begin{equation}
\label{Dsell}
P_\ell(k) = \frac{(-i)^\ell}{2\pi^2}\int_0^\infty dr\ r^2 j_\ell(kr)\ \xi_\ell(r).
\end{equation}

\noindent In this way, a 3D numerical integration gives both the redshift-space correlation function and power spectrum.


\begin{figure}[t!]
\begin{center}
\includegraphics[width=0.5\textwidth]{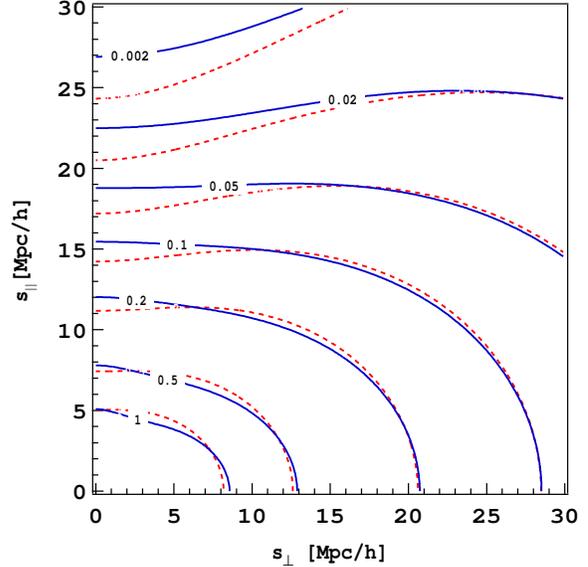}
\caption{Contours of $\xi_s(\spa,\spe)$ for the Exact Gaussian result (solid) and the Kaiser limit (dashed), for a flat $\Lambda$CDM cosmological model ($\Omega_m=0.26$, $\sigma_8=0.9$, $\Omega_b=0.04$, $h=0.7$) with linear bias $b_1=1$.}
\label{xis_Gau_Kai_a}
\end{center}
\end{figure} 

\begin{figure}[t!]
\begin{center}
\includegraphics[width=0.5\textwidth]{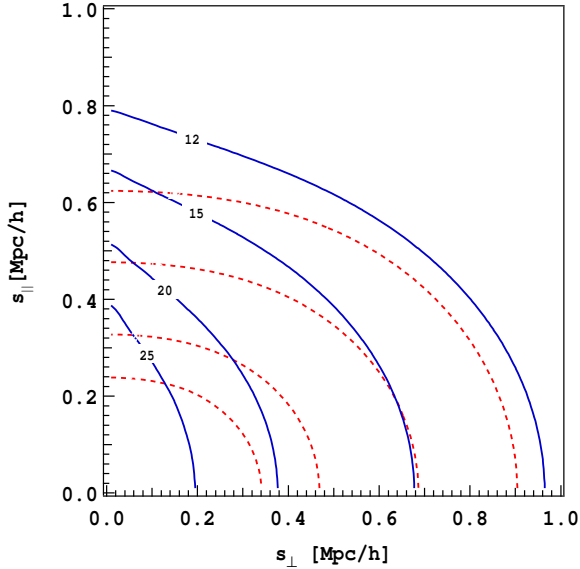}
\caption{Same as Fig.~\ref{xis_Gau_Kai_a} but at small scales. Note the dispersion effect: a Gaussian random field does show ``fingers of god"  even though the velocity dispersion decreases monotonically towards small scales (see dashed line in Fig.~\ref{v12s12}).}
\label{xis_Gau_Kai_b}
\end{center}
\end{figure} 



Figure~\ref{xis_Gau_Kai_a} shows the result for the redshift-space correlation function $\xi_s(\spa,\spe)$ in the exact Gaussian case Eq.~(\ref{xisexlin}) (solid) and the Kaiser limit (dashed), Eq.~(\ref{xislinder}) below. Notice that there are significant deviations even at large scales, predominantly at small $\spe$, we explain why this is the case in section~\ref{Pzlinls} below. Since the correlation amplitude is so much smaller in this region compared to small $\spa$, when multipoles are calculated integrating along fixed $s$  the results are close to their Kaiser limit values. It is apparent that the qualitative behavior of the corrections are to make the contours less squashed, as expected from the effects of the velocity dispersion. This is evident in Fig.~\ref{xis_Gau_Kai_b} which zooms into small scales, the dispersion effect is obvious (the quadrupole has opposite sign from that at large scales). Note that this happens at very small scales because the pairwise dispersion decreases at small scales (see dashed line in Fig.~\ref{v12s12}), thus one needs to go to tiny scales before $\spa$ becomes smaller than the pairwise dispersion. In addition, one can see that the Gaussian result is not close to the Kaiser limit even when the amplitude of the correlation function is much smaller than unity, e.g. see $\xi_s \simeq 0.002$ in the left panel in Fig.~\ref{xis_Gau_Kai_a}. 

As shown in Fig.~\ref{v12s12} and discussed above, assuming Gaussianity is not a good approximation, therefore these results are not a substantial improvement over the Kaiser limit, and we do not show corresponding results for the power spectrum. What is interesting here is that it gives some idea of how to incorporate the effects of large-scale velocity dispersion; we will come back later to this when we develop a simple approximation to the redshift-space power spectrum. We now  turn into a discussion of the assumptions behind the Kaiser limit, and show how our approach differs from the standard derivations of it in the literature.

\section{The Large-Scale Limit}
\label{Pzlinls}
\subsection{Derivation}
\label{deriv}

The non-trivial part of Eq.~(\ref{xisexlin}), and Eq.~(\ref{xiconv}) in general, is that as one integrates along $\rpa$ one is integrating over a different PDF due to scale dependence and anisotropy. The relationship between $\xi_s$ and $\xi$, $v_{12}$ and $\sigma_{12}^2$ ($\big\langle \Delta u_z^2 \big\rangle$ in linear dynamics) is non-local, however at large scales one can express $\xi_s$ in terms of local second moments by the following procedure, which leads to a derivation of the Kaiser formula and makes clear its regime of validity.

What do we exactly mean by ``large scales"?  Although $\xi$ and $v_{12}$ vanish in the large-scale limit,  $\sigma_{12}^2$ does not (and $\big\langle \Delta u_z^2 \big\rangle$ is {\em largest} at large scales), therefore we are not allowed to do a small amplitude expansion in this case. On the other hand, when $\spa \gg f\sigma_{12}$ the integration over $\rpa$ will be sharply peaked about $\rpa=\spa$, thus we can ``expand real space about redshift space",

\beqa
{\cal P}(v;\rpa) &\approx& {\cal P}(v;\spa) + (\rpa-\spa)\, \frac{d {\cal P}(v;\spa)}{d\spa} \nonumber \\
&+& \frac{1}{2} (\rpa-\spa)^2\, \frac{d^2 {\cal P}(v;\spa)}{d\spa^2}
+ \ldots
\label{raboutz}
\eeqa

Note that since this expansion can be done for any PDF (not just the one corresponding to linear dynamics), we will do so in general, our results here apply to the fully non-linear case. Similarly one can expand $\xi(r) \approx \xi(s) + \dots$, and using that $v=\rpa-\spa$ [see Eq.~(\ref{xiconv})], keeping up to second derivatives we obtain

\beqa
1+\xi_s &\approx& (1+\xi)\, \Big(1+ f v_{12}'  + 
\frac{f^2}{2} { \sigma_{12}^2}''+\ldots \Big) \nonumber \\
&&+\xi' f v_{12}+\frac{f^2}{2} \xi''  \sigma_{12}^2+ \xi' f^2 { \sigma_{12}^2}' + \ldots \nonumber \\ & &
\label{xisLSlimit}
\eeqa

\noindent where all quantities in the right hand side are evaluated at $\s$ and derivatives are with respect to $\spa$, e.g. $v_{12}' \equiv d v_{12}(\s)/d \spa$. Keeping only terms linear in quantities that vanish in the large-scale limit gives 

\beq
\xi_s \approx \xi + f v_{12}'  + 
\frac{f^2}{2} { \sigma_{12}^2}''+\frac{f^2}{2} \xi''  \sigma_{12}^2|_\infty,
\label{xisKlimit}
\eeq
where $\sigma_{12}^2|_\infty$ is the large-scale limit of the pairwise dispersion. In Fourier space this reads,

\beqa
P_s(\k) &\approx& P_{\d\d}(k)\, \Big(1-\frac{1}{2} f^2k_z^2\sigma_{12}^2|_\infty \Big)+ if k_z\, v_{12}(\k) 
\nonumber \\ & & - \frac{1}{2} f^2 k_z^2\, \sigma_{12}^2(\k).
\label{PsLSlimit}
\eeqa

\noindent Higher-order derivatives are suppressed by higher powers of $k_z$. Expanding real space about redshift space should work well when the derivatives in Eq.~(\ref{xisKlimit}) are small ($k_z$ is small), i.e. when considering waves with $\k$ with a small component with respect to the line of sight in which case the distortions are small. The large scale limit of $v_{12}$ is given by linear theory, $v_{12}(\k)=-2ik_z P_{\d\te}(k)/k^2$, whereas for $\sigma_{12}^2$ both Gaussian and non-Gaussian terms contribute. We calculate the non-Gaussian terms in paper II, for our purposes here let us just write $\sigma_{12}^2|_\infty=2(\sigma_v^2+A_\sigma)$ and $\sigma_{12}^2(\k)=-2k_z^2 P_{\te\te}(k)/k^4+B_\sigma(\k)$, then we have

\beqa
P_s(\k) &= & P_{\d\d}(k) \, \Big(1- f^2k_z^2(\sigma_v^2+A_\sigma) \Big)
+ 2f\, \frac{k_z^2}{k^2} \ P_{\d\te}(k)\nonumber \\ & & 
+ f^2\, \frac{k_z^4}{k^4} \ P_{\te\te}(k)- \frac{1}{2} f^2 k_z^2\, B_\sigma(\k),
\label{PLS}
\eeqa

\noindent where the non-Gaussian terms correspond to $A_\sigma= \langle u_z^2 \delta \rangle$ and $B_\sigma = {\rm FT} \langle \Delta u_z^2(\d+\d'+\d\d')\rangle_c$, where ${\rm FT}$ stands for Fourier transform. We show in paper II that in the large-scale limit, $B_\sigma \approx (8/35)(4+11\mu^2/3)\sigma_v^2 P(k)$. $A_\sigma$ corresponds to the difference in the large scale limit of $\sigma_{12}^2$ to the linear value (squares compared to dashed lines in Fig.~\ref{v12s12}), whereas $B_\sigma$ is the non-Gaussian contribution that takes into account that the scale dependence of the pairwise velocities is {\em opposite} to that in linear theory, i.e. increasing toward smaller scales, as a result it counters the effect of the $P_{\te\te}$ term.  From Eq.~(\ref{PLS}) it follows that when 

\beq
\label{kzlimit}
k_z^2\ f^2 \sigma_v^2 \ll 1, \ \ \ \ \ \ \ \ \ \ {\rm or}\ \ \ \ \ \ \ \ \ \ 
k \mu \ll 0.2~h~{\rm Mpc^{-1}},
\eeq 

\noindent where we assumed a flat $\Lambda$CDM model, for which $ \sigma_v^2 \approx 40 $ (Mpc $h^{-1}$)$^2$ and $f \approx 0.5$ at $z=0$, one recovers the Kaiser formula~\cite{1987MNRAS.227....1K} for the power spectrum (the reason why we don't assume $P_{\d\d}=P_{\d\te}=P_{\te\te}$ will become clear in section~\ref{theta1L}),

\beq
P_s(\k) = P_{\d\d}(k)+ 2f\mu^2 P_{\d\te}(k)+ f^2\mu^4 P_{\te\te}(k).
\label{PKaiser}
\eeq

The condition in Eq.~(\ref{kzlimit}) says that, unless one considers modes nearly perpendicular to the line of sight $\mu \sim 0$, velocity dispersion effects become important for wavenumbers much smaller than the non-linear scale. Note that at $k_z \sim 0.2~h~{\rm Mpc^{-1}}$ the velocity dispersion terms become {\em of order unity} almost reversing the enhancement of the redshift-space power spectrum. These additional terms have important dependencies on cosmological parameters that are different from those in the Kaiser formula, for example $A_\sigma \sim b_1 \sigma_8^2$ and $B_\sigma\sim b_1 \sigma_8^4$ in the large scale limit, where $b_1$ is the linear bias, with $\sigma_v\sim\sigma_8$ depending also on the shape of the power spectrum. This can help break degeneracies present in Eq.~(\ref{PKaiser}). 

Note that although Eq.~(\ref{PKaiser}) has the right limit at $k_z=0$, giving the real-space power spectrum, the second derivative (which is the first non-vanishing) with respect to $k_z$ does not (except at $k=0$), as this is sensitive to velocity dispersion effects, both Gaussian and non-Gaussian. It is useful to recast Eq.~(\ref{PKaiser}) in terms of what it implies for the pairwise velocity PDF. To do this, we can expand  Eq.~(\ref{Zlinear}) for small $\lambda$ (recall $\lambda=ifk_z$ in Fourier space),

\begin{equation}
\label{Zapprox}
{\cal Z}_{\rm G} \approx 1+\xi(r) + \lambda\ v_{12}(r) + \frac{\lambda^2}{2}\ 
\big\langle \Delta u_z^2 \big\rangle.
\end{equation}
This implies that the pairwise velocity PDF in the Kaiser limit has the form [see Eq.~(\ref{Pv})],

\begin{equation}
\label{PvK}
{\cal P}(v) \approx \Bigg(1-f\ v_{12}\ \frac{d}{d v} +\frac{f^2}{2} \big\langle \Delta u_z^2 \big\rangle \ \frac{d^2}{d v^2} \Bigg)\ \delta_D(v),
\end{equation}

\noindent that is, it corresponds to a very sharply peaked PDF, since the dispersion $\big\langle \Delta u_z^2 \big\rangle$ is effectively assumed to be vanishingly small. This is the result used in Eq.~(\ref{Pvdisp}) to derive the pairwise PDF in the dispersion model, and when put into Eq.~(\ref{xiconv}) gives the two-point function~\cite{1992ApJ...385L...5H,1995ApJ...448..494F}) 

\beq
\label{xislinder}
\xi_s(\spa,\spe)=\xi(s)+f\frac{d}{d\spa} v_{12}(\s)+\frac{f^2}{2}\frac{d^2}{d\spa^2}\langle\Delta\u^2\rangle.
\eeq

It is interesting to go back to Fig.~\ref{xis_Gau_Kai_a} and compare the exact result for Gaussian random fields to the Kaiser formula. The expansion in Eq.~(\ref{raboutz}) is best when the scale dependence of the PDF is small.  This is going to be less safe for smaller $\spe$, since for large $\spe$ variations in $\rpa$ as one integrates enter only in quadrature in $s^2=\rpa^2+\spe^2$, whereas for $\spe \simeq 0$ variations in $\rpa$ enter linearly into $s$. This is the analogous situation to having $k_z$ not small in Fourier space, and this is why the largest deviations in Fig.~\ref{xis_Gau_Kai_a} happen near $\spe=0$, even at large scales.

Finally, a few words of caution about the expansion in Eq.~(\ref{raboutz}). This converts integration over an infinite number of PDF's into a single one and its derivatives, thus significantly simplifying the calculation. Note however than in order to arrive to Eq.~(\ref{xisKlimit}) one must interchange the order of the derivatives and integrals over the PDF and integrate term by term. Such a procedure is not strictly valid, since it is very likely that the expansion  in Eq.~(\ref{raboutz}) does not converge uniformly.  Indeed, in the Gaussian case one is expanding Eq.~(\ref{Zlinear}) for small $\lambda$, and the exponential series has zero radius of convergence, thus term by term integration is not mathematically valid. Note also that at the end, terms that were supposed to be of increasing order in a small parameter in Eq.~(\ref{raboutz}) end up being of the same order of magnitude in Eq.~(\ref{PKaiser}).

\subsection{Comparison with the standard derivation}

Let us now compare our derivation of the Kaiser limit with the standard approach (\cite{1987MNRAS.227....1K,1994MNRAS.267..785C,1998evun.work..185H}) which makes explicit use of the Jacobian $J= |\partial s_i/\partial x_j|$ of the mapping from real to redshift space. In the plane-parallel approximation, $J(\x)=|1-f\nabla_z \u|$ and from Eq.~(\ref{mass}) it follows that $1+\d_s(\s)=[1+\d(\x)]/J(\x)$. Now {\em if we assume} $f\nabla_z \u \ll 1$, we can expand $1/J(\x) \simeq 1+f\nabla_z \u$, and thus linearizing in the {\em field} amplitudes if follows,

\beq
\d_s(\s) \simeq \d(\x) +f\nabla_z \u(\x),
\label{dss}
\eeq
which, using that $\nabla \cdot {\bf u}=\d$ in linear dynamics, and $\s \simeq \x$ to leading order, in Fourier space leads to $\d_s(\k)=\d(\k) (1+f \mu^2)$.  

There are several steps in this derivation which are unjustified, namely, the density and velocity gradients  {\em at a given point in space} are not small (i.e. for CDM models their linear variance at a point  is much larger than unity), note that there is no smoothing involved until {\em after} one makes these approximations. In particular $\d(\x)$ can be large inside dark matter halos and similarly $\nabla_z \u$, which will also fluctuate in sign. What is small is the {\em correlation} between fields separated by large distances, not the field amplitudes themselves. By making approximations at the level of density and velocity fields one gets incorrect correlations, in the sense that the velocity dispersion of a Gaussian random field never appears  in this approach. The derivation presented in section~\ref{sec21} and~\ref{deriv} shows that it is unnecessary to assume anything about the Jacobian of the transformation or the amplitude of density and velocity gradients.

\section{Non-linear Evolution of Density and Velocity Fields}
\label{theta1L}

The expansion leading to Eq.~(\ref{PLS}) has little to do with nonlinear dynamics (only involved in generating the non-Gaussian terms), but rather with the nonlinearities of the real to redshift-space mapping. We now explore the corrections induced in the redshift-space power spectrum due to non-linear evolution of the density and velocity fields. We shall see that the velocity field is affected more significantly than the density field at large scales due to larger sensitivity to tidal gravitational fields.  

We are interested in calculating the non-linear evolution of density and velocity divergence auto and cross spectra and comparing to numerical simulations.  Measuring the {\em volume-weighted} velocity divergence power spectrum in numerical simulations is not straightforward at small scales. Interpolating the particles velocities to a grid gives the {\em momentum} (density-weighted velocities); in order to obtain $\theta(\k)$ one possibility would be to

\begin{enumerate}
\item[i)] Fourier transform the momentum, and divide the Fourier coefficients by the interpolation window (``sharpening" of the momentum Fourier coefficients).

\item[ii)] do the same for the density field, and then transform back to real space density and momentum fields.

\item[iii)] divide momentum by density at each grid point. Fourier transform the resulting volume-weighted velocity field and calculate the divergence in Fourier space.

\end{enumerate}

This procedure is not ideal for several reasons. First, there is the choice of the interpolation scheme: one would like to choose a low-order interpolation scheme because it does not smooth out fields too much (so sharpening only affects the highest-$k$ modes), on the other hand, a low-order interpolation scheme gives rise to many grid points with zero density and momentum, thus the velocity field cannot be defined there. Using a high-order interpolation scheme bypasses this problem, but leads to some grid points with negative density after step ii), due to the fact that sharpening can be numerically unstable in voids. A more practical procedure is to divide the interpolated momentum by the interpolated density (both of which have been similarly affected by the interpolation window), Fourier transform that, and without applying any corrections (since interpolation corrections in numerator and denominator should roughly cancel), calculate the divergence of the velocity field. This procedure is safe to the extent that gives results independent of the interpolation scheme. We have tried second (CIC), third (TSC) and fourth-order interpolation schemes with similar results: at large scales $k \la 0.3~h$~Mpc$^{-1}$ the different procedures give the same power spectrum, for smaller scales the results obtained start to depend on the particular scheme used. It would be interesting to try using Delaunay or Voronoi tesselation techniques~\cite{1996MNRAS.279..693B} to see whether this can be improved for smaller scales, but our procedure is simpler and works well at large scales.

We now present the calculation of the density and velocity divergence auto and cross power spectra using one-loop PT. In linear PT, by definition $P_{\d\d}(k)=P_{\d\te}(k)=P_{\te\te}(k)\equiv P(k)$. Non-linear corrections break this degeneracy, giving 

\begin{widetext}
\beqa
P_{\d\d}(k)&=& P(k)+ 2 \int [ F_2(\p,\q)]^2 P(p) P(q) d^3q +
6 P(k) \int F_3(k,q) P(q) d^3q \\
P_{\te\te}(k)&=& P(k)+ 2 \int [ G_2(\p,\q) ]^2 P(p) P(q) d^3q +
6 P(k) \int G_3(k,q) P(q) d^3q \\
P_{\d\te}(k)&=& P(k)+ 2 \int F_2(\p,\q) G_2(\p,\q) P(p) P(q) d^3q + 
 3 P(k) \int [F_3(k,q)+G_3(k,q)] P(q) d^3q,
\eeqa
\end{widetext}

\noindent where $\p=\k-\q$. The first term of non-linear corrections describes the contribution to the power spectrum at $\k$ due coupling between modes $\q$ and $\p$, whereas the second term corresponds instead to corrections to the linear growth factor that depend on $k$. The kernels $F_2$ and $G_2$ can be written as ($\tvk=\k/k$ and similarly for $\tvq$)

\begin{widetext}
\beqa
F_2(\k,\q) &=& \frac{\nu_2}{2} +\frac{1}{2}\, \tvk \cdot \tvq \, \Big(\frac{k}{q}+\frac{q}{k}\Big) +
\frac{2}{7}\, \Big(\tvk_i \tvk_j -\frac{1}{3}\d_{ij}\Big)\Big(\tvq_i \tvq_j -\frac{1}{3}\d_{ij}\Big)
 \\ & & \nonumber \\
G_2(\k,\q) &=& \frac{\mu_2}{2} +\frac{1}{2}\, \tvk \cdot \tvq \, \Big(\frac{k}{q}+\frac{q}{k}\Big) +
\frac{4}{7}\, \Big(\tvk_i \tvk_j -\frac{1}{3}\d_{ij}\Big)\Big(\tvq_i \tvq_j -\frac{1}{3}\d_{ij}\Big)
\eeqa
\end{widetext}
where $\nu_2=34/21$ and $\mu_2=26/21$ represent the second-order evolution in the spherical collapse dynamics. The other two terms in these kernels have a different physical origin: the middle term is due to the non-linear transformation from following mass elements to studying the dynamics at fixed spatial position (``Lagrangian to Eulerian space" mapping), the last term represents the effect of the tidal gravitational fields, since $(\tvk_i \tvk_j -\frac{1}{3}\d_{ij})\d(\k)$ is the Fourier representation of the tidal gravitational field $\nabla_i\nabla_j\Phi(\x)-\frac{1}{3}\d_{ij}\nabla^2\Phi(\x)$, where $\Phi$ is the gravitational potential. The important thing to notice here is that velocity fields are more sensitive to tidal fields, the coefficient of the last term in $G_2$ is twice that in $F_2$, and consequently they evolve less by spherical collapse (that's why $\mu_2$ is smaller than $\nu_2$ to exactly compensate) and therefore do not grow as fast due to non-linear effects, in fact, we shall see that non-linear growth is significantly {\em smaller} than linear at the scales we are interested in. 

The $F_3$ and $G_3$ terms can be analyzed in a similar way, but they are more complicated, instead we just write down their expression after the angular integration over $\tvk \cdot \tvq$ has been done,

\beqa
F_3(k,q)&=&\frac{6k^6-79k^4q^2+50k^2q^4-21q^6}{63k^2q^4} \nonumber \\ &+&
\frac{(q^2-k^2)^3 (7q^2+2k^2)}{42k^3q^5}\ \ln \Big|\frac{k+q}{k-q}\Big| , \nonumber \\ 
& &  \\
G_3(k,q)&=&\frac{6k^6-41k^4q^2+2k^2q^4-3q^6}{21k^2q^4}  \nonumber \\ &+&
\frac{(q^2-k^2)^3 (q^2+2k^2)}{14k^3q^5}\ \ln \Big|\frac{k+q}{k-q}\Big| .\nonumber \\ 
& &
\eeqa

\begin{figure}[t!]
\begin{center}
\includegraphics[width=0.5\textwidth]{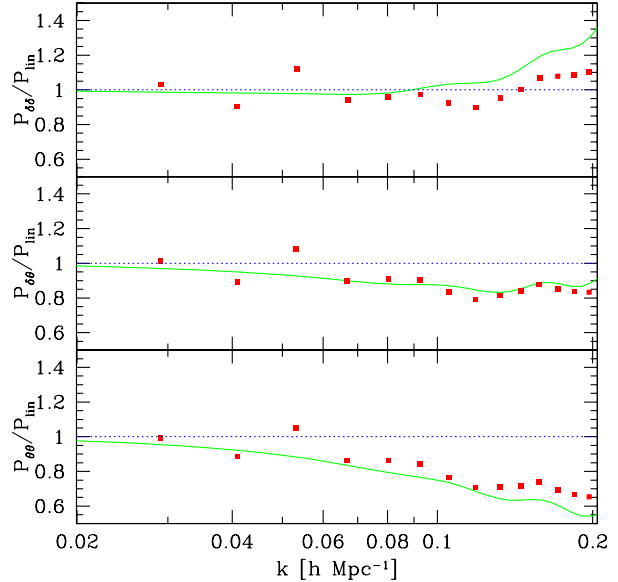}
\caption{Non-linear corrections to the density-density (top), density-velocity (middle) and velocity-velocity (bottom) power spectra as a function of scale. The symbols denote measurements in the VLS dark matter simulations, solid lines denote one-loop perturbation theory.}
\label{Pk_dd_dv_vv_vls512}
\end{center}
\end{figure}

These terms are negative and the magnitude of $G_3$ is larger than $F_3$. This leads to an overall suppression of $P_{\te\te}(k)$ compared to linear theory. Figure~\ref{Pk_dd_dv_vv_vls512} shows the results of these calculations (solid lines) and measurements in numerical simulations (symbols), expressed as ratios to the linear power spectrum $P_{\rm lin}(k)$.  One-loop PT for $P_{\d\d}(k)$ performs significantly worse than for spectra with no baryonic wiggles, though it does seem to track the variations seen in the simulations, about $10\%$ for $k\la 0.2~h$~Mpc$^{-1}$, at least in a qualitative sense. The situation is significantly better for $P_{\d\te}(k)$, but this good agreement appears to be to some extent an accident, a cancellation between too large corrections for $P_{\d\d}(k)$ and $P_{\te\te}(k)$ with opposite signs.

These results can be understood qualitatively and to some extent quantitatively as well by considering one-loop PT for scale-free initial conditions with $P_{\rm lin}(k) =k^n$. In this case~\cite{1996ApJ...473..620S},

\beq
P_{xy}(k)=P_{\rm lin}(k)\ \Big[ 1+ \alpha_{xy}(n)\, \Big(\frac{k}{k_{\rm nl}}\Big)^{n+3} \Big],
\eeq

\noindent where $k_{\rm nl}$ is the nonlinear scale defined form the linear power spectrum, and $x$ and $y$ denote any of $\d,\te$. The functions $\alpha$ are decreasing functions of $n$, positive for $n$ sufficiently negative and negative for $n$ sufficiently positive; the sign of $\alpha$ describes whether 
 the nonlinear growth is faster or slower than in linear theory. Corrections to $P_{\d\d}$ ($P_{\te\te}$) are positive for $n<-1.4$ ($n<-1.9$) and negative otherwise (see Fig.~12 in~\cite{2002PhR...367....1B} for plots of $\alpha_{\d\d}$ and $\alpha_{\te\te}$). For CDM models close to the nonlinear scale at e.g. $k=0.1~h$~Mpc$^{-1}$, the effective spectral index is $n_{\rm eff} \approx -1.35$, which being close to the critical index for $\d$ where corrections to $P_{\d\d}(k)$ vanish, leads to a small negative correction to $P_{\d\d}$. On the other hand, the situation is very different for $\te$ that has a critical index of $-1.9$, thus the large negative corrections to $P_{\te\te}$~\footnote{A similar (reversed) situation happens at high redshift, where the spectral index close to the nonlinear scale becomes very negative. In this case corrections to $P_{\d\d}$ become large and positive (see e.g. Fig.~3 in~\cite{2003ApJ...590....1Z}). This is another example where the meaning of a ``nonlinear scale" from the linear power spectrum can be very misleading. Due to tidal effects the growth of density perturbations is not just a function of the amplitude of the density field at a given point (as in the spherical collapse model) but also has important dependence on the shape of the power spectrum.}. This has a significant impact on the large-scale redshift-space power spectrum. For more discussion of nonlinear corrections along these lines see e.g.~\cite{1992PhRvD..46..585M,1996ApJ...467....1L,1996ApJ...473..620S,1998MNRAS.301..535F,2002PhR...367....1B}. 

\section{The redshift-space power spectrum}
\label{rsps}

\begin{figure}[t!]
\begin{center}
\includegraphics[width=0.5\textwidth]{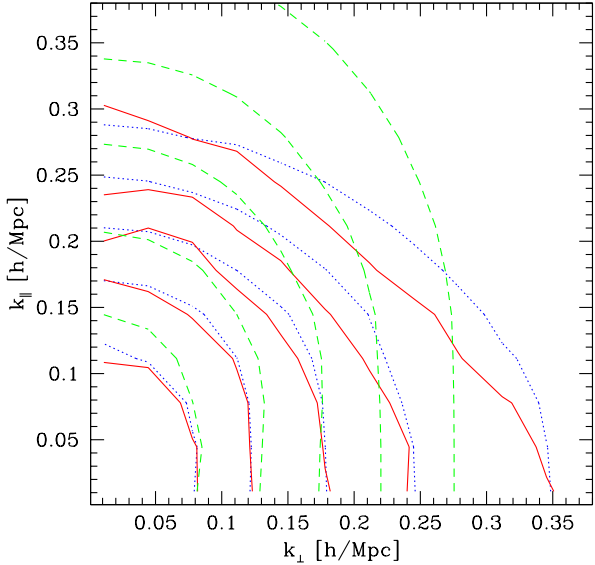}
\caption{Contours of the redshift-space power spectrum at $z=0$. The solid lines correspond to the N-body simulation results, dashed lines denote the Kaiser formula, and dotted lines  show the simplified ansatz of Eq.~(\ref{PkA}).}
\label{Pk_contours}
\end{center}
\end{figure}

\subsection{A simple model}
\label{simpmod}

We now put together the results discussed above to see how well one can match the large-scale redshift-space power spectrum with the information we have so far, without resorting to an evaluation of the PDF of pairwise velocities in the non-Gaussian case. Specifically, we use the following ansatz,

\beqa
P_s(\k) &=& \Big( P_{\d\d}(k)
+ 2f\mu^2 \, P_{\d\te}(k)+ f^2\mu^4 \, P_{\te\te}(k)\Big) \nonumber \\
& & \times\ {\rm exp}(-f^2k_z^2 \sigma_v^2),
\label{PkA}
\eeqa
where $P_{\d\d}$, $P_{\d\te}$ and $P_{\te\te}$ refer to the {\em non-linear} spectra, see Fig.~\ref{Pk_dd_dv_vv_vls512}. We only include velocity dispersion effects using the large-scale limit in the Gaussian case; as discussed at the end of section~\ref{PmomG} this {\em is not} correct even at large scales, as the pairwise velocity PDF is significantly non-Gaussian at all scales. Going beyond this however requires and evaluation of the pairwise PDF in the non-Gaussian case, which is addressed  in paper II. We try to compensate for this by keeping a {\em constant} Gaussian velocity dispersion suppression factor given by linear dynamics; this is an improvement over the incorrect scale dependence in linear dynamics and partially mimics the effect of non-Gaussian terms. But it is clearly an oversimplification. Note that although at first sight Eq.~(\ref{PkA}) looks similar to the phenomenological model of~\cite{1994MNRAS.267.1020P}, it is in fact rather different: we do not fit for a velocity dispersion factor, but rather $\sigma_v^2$ is predicted by linear dynamics and depends on $\sigma_8$ and the shape of the power spectrum; also, we incorporate the difference in evolution between density and velocity fields at large scales, as seen in Fig.~\ref{Pk_dd_dv_vv_vls512}.

Figure~\ref{Pk_contours} shows the results of such an exercise, compared to the numerical simulation results (symbols) and to the Kaiser formula (dashed). Although the improvement is significant there are still some deviations, which is not surprising given our approximate treatment.  In particular, Eq.~(\ref{PkA}) does not give enough suppression at intermediate angles. The suppression of power at $\mu=1$ works reasonably well, and it is due to the velocity dispersion {\em and} the nonlinear corrections to $P_{\d\te}$ and $P_{\te\te}$; for example at $k=0.1~h$~Mpc$^{-1}$ each effects suppresses power at $\mu=1$ by the same amount, about $10\%$ each. 

\subsection{Recovering the real-space power spectrum}

\begin{figure}[t!]
\begin{center}
\includegraphics[width=0.5\textwidth]{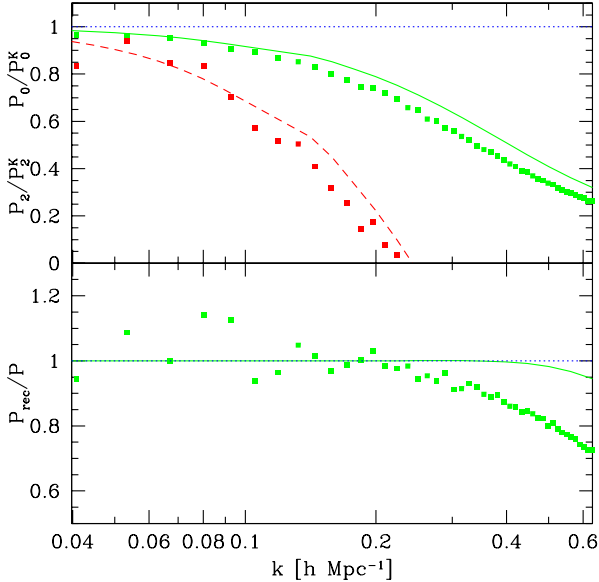}
\caption{{\em Top panel:} Ratio of the monopole (solid) and quadrupole (dashed) to the predictions of the Kaiser formula, for redshit-space power given by Eq.~(\ref{PkA}). Symbols show the same quantities in the numerical simulations. {\em Bottom panel:} recovery of the real-space power spectrum from redshift-space multipoles according to Eq.~(\ref{reconPk}) for the model in Eq.~(\ref{PkA}) (solid) and N-body simulations (symbols).}
\label{P0P2recov}
\end{center}
\end{figure}

An important question is to what extent one can recover the real-space power from measurements of the redshift-space power spectrum. Attempts to do this fall into two different approaches: one is to measure the projected correlation function $\xi_p$, Eq.~(\ref{xip}), by integrating the redshift-space correlation function along the line of sight~\cite{2001MNRAS.328...64N,2002ApJ...571..172Z,2002ApJ...564...15J}, the other is to try to measure the power for modes perpendicular to the line of sight either by smoothly approaching $\mu=0$~\cite{2002MNRAS.330..506H} or at large scales by using the Kaiser formula to go from multipoles to the real-space power~\cite{2002MNRAS.335..887T,2004ApJ...606..702T}. Here we explore the conditions of validity of the latter. 

First, it is important to note that even if one is interested in just the shape of the real-space power spectrum (and not its amplitude), it should be clear from the results presented so far that there is no reason to expect that the monopole of the redshift-space power should have the same shape as the real-space power, at least not to the accuracy of current large surveys such as 2DFGRS and SDSS. The top panel in Fig.~\ref{P0P2recov} illustrates this point for the model of Eq.~(\ref{PkA}) and N-body simulations, where we compare their monopole and quadrupole to those in the Kaiser limit, $P^{\rm K}=P(k) (1+f\mu^2)^2$ with $P(k)$ the nonlinear real-space power spectrum. Note that even at $k=0.1~h$~Mpc$^{-1}$ the monopole is suppressed by about $10\%$ and the quadrupole by $35-40\%$. From Fig.~\ref{P0P2recov} we can see again that our model underestimates the suppression when compared to numerical simulations. In principle the situation for galaxies could be different than shown in Fig.~\ref{P0P2recov}, but for close to unbiased galaxies there is no reason why it should be drastically different than for the model in Eq.~(\ref{PkA}), given that we only include velocity dispersion due to large-scale flows and nonlinear corrections to $\theta$ should not be affected by biasing, being a volume weighted velocity [see discussion after Eq.~(\ref{Zgen2})]. We stress that ignoring the suppression of power at large scales can contribute to systematic effects in the determination of shape parameter, the spectral tilt and running of the spectral index, or constraints on the neutrino mass.

The good news is that an ``inverse use" of the Kaiser formula has a larger regime of validity than one might expect based on the results discussed so far. As long as we can approximate the redshift-space power spectrum {\em with only}  $\ell=0,2,4$ multipoles we can always write

\beq
P_s(\k) \equiv P(k)\ [1+2A_2(k)\, \mu^2+A_4(k)\, \mu^4]
\label{L024app}
\eeq
with $A_2(k)$ and $A_4(k)$ some arbitrary functions of $k$. One can think of these functions as scale dependent versions of $f$ or $\beta$ when bias is present, i.e. in the Kaiser limit  $A_4(k)=[A_2(k)]^2=\beta^2$.  The interesting piece of information is that recovering the real-space power spectrum from the redshift-space multipoles in the case of arbitrary $A_2$ and $A_4$ is that is still given by the {\em same} linear combination as in the Kaiser limit, 

\beq
P(k)=P_0(k)-\frac{1}{2}P_2(k)+\frac{3}{8}P_4(k),
\label{reconPk}
\eeq 

\noindent even for {\em arbirtrary} $A_2(k)$ and $A_4(k)$, since Eq.~(\ref{reconPk}) only uses orthogonality of Legendre polynomials up to $\ell=4$. Equation~(\ref{reconPk}) is thus far more general than assuming the Kaiser limit, basically the linear combination at each scale is done using the effective value of $\beta$ at that scale implied by $A_2$ and $A_4$. The reason why this is useful is that higher than $\ell=4$ multipoles are generated only for $k\ga0.2~h$~Mpc$^{-1}$ since they are suppressed by higher powers of $k_z$ in the large-scale expansion, see Eq.~(\ref{raboutz}). An example of the effectiveness of using Eq.~(\ref{reconPk}) is given in the bottom panel of Fig.~\ref{P0P2recov}, where we use it to reconstruct the real power in the case of the model in Eq.~(\ref{PkA}) and for the N-body measurements, which {\em do not} have the form of Eq.~(\ref{L024app}), since the exponential generates all multipoles higher than $\ell=4$ with roughly equal amplitude in the high-$k$ limit and even more so for the simulation. Nonetheless, the recovery of the real-space power is quite successful for $k\la0.2-0.3~h$~Mpc$^{-1}$, a bit worse for the simulation that has a larger velocity dispersion everywhere compared to the Gaussian value (see top panel in Fig.~\ref{v12s12}).

The approach of using Eq.~(\ref{reconPk}) to recover the real-space power spectrum was implemented already in~\cite{2002MNRAS.335..887T,2004ApJ...606..702T}.  Of course the use Eq.~(\ref{reconPk}) can be extended to include higher multipoles if possible, this will increase the regime of validity of the reconstruction. Note however that the nice property of recovering $P_{\d\d}$ does not extend to $P_{\d\te}$ and $P_{\te\te}$, e.g. using the same idea one obtains $\frac{3}{4}P_2(k)-\frac{15}{8}P_4(k)=A_2(k) P(k)$, which cannot be interpreted as $P_{\d\te}$ with the same degree of accuracy due to the effects of velocity dispersion.

\section{Conclusions}
\label{concl}

We have derived the exact relationship between two-point statistics in real and redshift space in terms of the statistics of pairwise velocities. This is given by Eq.~(\ref{xiconv}) for the two-point correlation function in terms of the pairwise velocity PDF, and by Eq.(\ref{powerz}) for the power spectrum in terms of the pairwise velocity generating function. These results include all non-linearities in the dynamics and the real-to-redshift space mapping, the only approximation made is that distortions are plane-parallel. The radial distortion case can be derived by similar reasoning to that in section~\ref{sec21}. Higher-order correlation functions in redshift space can also be studied along the same lines. 

We also showed that,

\begin{enumerate}

\item[i)] The pairwise velocity PDF is strongly non-Gaussian at all scales (Figs.~\ref{PDFpar} and~\ref{PDFperp}). The failure to reach Gaussianity at large scales is related to the fact that difference of velocities between members of a pair are always sensitive to modes whose wavelength is smaller than the distance of separation. 

\item[ii)] The often used dispersion model, Eq.~(\ref{DMexp}), gives rise to an unphysical distribution of  pairwise velocities (see bottom left panel in Fig.~\ref{PDFpar}).

\item[iii)] It is impossible in general to derive the PDF of pairwise velocities from measurements of redshift-space clustering. Methods that claim to do this obtain instead something else, whose properties we derive, see Eqs.~(\ref{m2w}-\ref{sigeff}). 

\item[iv)] The exact result for the redshift-space correlation function of a random Gaussian field is significantly different from the Kaiser formula at large scales for pairs parallel to the line of sight (Fig.~\ref{xis_Gau_Kai_a}).

\item[v)] The large-scale limit of the redshift-space power spectrum in the general case differs from the Kaiser formula by terms that depend on Gaussian and non-Gaussian contributions to the velocity dispersion of large-scale flows, Eq.~(\ref{PsLSlimit}). 

\item[vi)] There are significant nonlinear corrections to the evolution of velocity fields at scales much larger than the ``nonlinear scale" (Fig.~\ref{Pk_dd_dv_vv_vls512}). These are due to the sensitivity of velocities to tidal gravitational fields, which suppress the growth relative to linear perturbation theory and have a significant impact on the redshift-space power spectrum. These corrections should be included when modeling large-scale velocity flows. 

\item[vii)] The monopole of the redshift-space power spectrum does not provide a good measure of the shape of the real-space power spectrum (top panel in Fig.~\ref{P0P2recov}). Ignoring this can lead to systematic effects in the determination of the spectral tilt, running of the spectral index, and limits on the neutrino mass.

\item[viii)] The real-space power spectrum can be recovered at large scales by the standard procedure based on the orthogonality of multipoles (bottom panel in Fig.~\ref{P0P2recov}).

\end{enumerate}

We have ignored the problem of galaxy biasing, although linear bias is of course trivial to introduce; it is interesting to note in this regard that non-Gaussian terms give a different dependence on cosmological parameters that can be used to break degeneracies. Nonlinearities in the bias between galaxies and dark matter can lead to nontrivial behavior, this will be explored elsewhere. An important gap that remains is the derivation of the large-scale limit of the PDF of pairwise velocities, this is a difficult problem that will be addressed in paper II. This should allow a more physical modeling of the redshift-space power spectrum along the lines of section~\ref{simpmod}, where we assumed (incorrectly) that Gaussianity holds at large scales. The usefulness of such a model is that it allows for the correlation that exists between the squashing and dispersion effects, which so far have been taken as independent in the modeling of redshift distortions, such as Eq.~(\ref{DMexp}). The correlation between both effects depends on $\Omega_m$, and the shape and normalization of the power spectrum. Using this information is essential to extract the full information encoded in the anisotropy of the redshift-space power spectrum, which on physical grounds must be poorly described by just two independent numbers such as $\beta$ and an effective velocity dispersion $\sigma_p$.

\acknowledgments

We thank Andreas Berlind,  Mart\'{\i}n Crocce, Josh Frieman, Enrique Gazta\~naga,  Roman Juszkiewicz, Dmitry Pogosyan, Alex Szalay, Max Tegmark, Jeremy Tinker, and David Weinberg for useful discussions. I benefited greatly from feedback  and innumerable discussions with Andrew Hamilton, Lam Hui and Ravi  Sheth. Some of this work has been carried out while at the Aspen Center for Physics.  We thank Naoki Yoshida for help dealing with the VLS simulations, which were carried out by the Virgo Supercomputing Consortium using computers based at the Computing Centre of the Max-Planck Society in Garching and at the Edinburgh parallel Computing Centre. The data are publicly available at {\tt http://www.mpa-garching.mpg.de/NumCos}.


\bibliography{masterbiblio}

\end{document}